\documentclass[notitlepage,reprint,onecolumn, amsmath,amssymb, aps,prd, nofootinbib]{revtex4-1} 

\usepackage{graphicx}
\usepackage{axodraw2}
\usepackage{epsfig}

\usepackage{bm}
\usepackage{color}
\usepackage[utf8]{inputenc} 
\usepackage[T1]{fontenc}

\usepackage{xfrac}
\usepackage{booktabs} 
\usepackage{array} 
\usepackage{paralist} 
\usepackage{verbatim} 
\usepackage{subfig} 
\usepackage{braket}
\usepackage{slashed}

\numberwithin{equation}{section}

\def\bit{\begin{itemize}}
\def\eit{\end{itemize}}
\def\ben{\begin{enumerate}}
\def\een{\end{enumerate}}
\def\beq{\begin{equation}}
\def\eeq{\end{equation}}
\def\bea{\begin{eqnarray}}
\def\eea{\end{eqnarray}}
\def\bq{\begin{quote}}
\def\eq{\end{quote}}
\def \lsim{\mathrel{\vcenter
     {\hbox{$<$}\nointerlineskip\hbox{$\sim$}}}}
\def \gsim{\mathrel{\vcenter
     {\hbox{$>$}\nointerlineskip\hbox{$\sim$}}}}
\def\gappeq{\mathrel{\rlap {\raise.5ex\hbox{$>$}}
{\lower.5ex\hbox{$\sim$}}}}
\def\lappeq{\mathrel{\rlap{\raise.5ex\hbox{$<$}}
{\lower.5ex\hbox{$\sim$}}}}

\def\Dslash{ \, D  \! \! \! \! / ~ }

\def\mec{\mu \! \to \! e~ {\rm conversion}}

\def\meee{\mu \to e \bar{e} e}

\def\tlp{\tau \to \ell \pi^0 }
\def\tlr{\tau \to \ell \rho }
\def\tnp{\tau \to \nu \pi^- }

\def\tlpp0{\tau \to \ell \pi^0 \pi^0}
\def\tlpp{\tau \to \ell \pi^+ \pi^-}
\def\tnpp{\tau \to \nu \pi^0 \pi^-}

\def\a{\alpha}
\def\b{\beta}
\def\g{\gamma}
\def\d{\delta}

\def\r{\rho}
\def\s{\sigma}

\begin{document}

\title{Charged lepton flavour change and Non-Standard neutrino  Interactions}

\author{Sacha Davidson}
\email{E-mail address: s.davidson@lupm.in2p3.fr}
\affiliation{LUPM, CNRS,
Université Montpellier
Place Eugene Bataillon, F-34095 Montpellier, Cedex 5, France
}
\author{Martin Gorbahn}
\email{E-mail address: martin.gorbahn@liverpool.ac.uk}
\affiliation{Theoretical Physics Division,
Department of Mathematical Sciences,
University of Liverpool,\\
 Liverpool L69 3BX, United Kingdom
}

\begin{abstract}
\noindent

Non-Standard neutrino Interactions  (NSI)
are  vector contact interactions involving
two neutrinos and two first generation fermions, which can affect
neutrino propagation in matter. 
 SU(2) gauge invariance suggests that NSI 
 should be accompanied by more observable  charged lepton contact interactions.
 However, these can be avoided at tree level in various
 ways. 
We focus on lepton flavour-changing NSI,  suppose they are generated
by New Physics heavier than $m_W$ that does not induce (charged)  Lepton
Flavour Violation (LFV) at tree level, and
show that LFV is generated 
at one loop  in  most cases.
The current constraints on charged Lepton
Flavour Violation therefore suggest that
$\mu \leftrightarrow e$ flavour-changing NSI are unobservable
and $\tau \leftrightarrow \ell$ flavour-changing NSI are
an order of magnitude weaker than the weak interactions.
This conclusion can be avoided if
the heavy New Physics conspires to
cancel the one-loop LFV,
or if NSI are generated by light
New Physics to which our analysis does not apply.

\end{abstract}

\maketitle


\section{Introduction and review}
\label{intro}

  Non-Standard neutrino Interactions (NSI)
  are  four-fermion interactions induced by physics
  from  Beyond-the-Standard Model,
  constructed from  a vector current of  two  Standard Model (SM) neutrinos
 of flavour $\rho$ and $\s$,
  and two  first generation fermions $f \in \{ e,u,d\}$.
  Below the weak scale, such  interactions
  can be  included in the Lagrangian  as
\beq
- 2\sqrt{2} G_F  \varepsilon_f^{\r\s}(\overline{\nu} _\r \g_\a  P_L \nu_\s)
(\overline{f}\g^\a P_X f )\label{eqn1}
\eeq
where  $G_F= 1/(2\sqrt{2} v^2)$ is the Fermi constant, 
the  dimensionless  coefficient $\varepsilon^{\r\s}$
parametrises the  strength of these new  interactions,
 $P_X$ is a chiral projector $P_{L/R} = (1\pm \g_5)/2$,
and $f$ will  be referred to  as the ``external'' fermion.

NSI  were introduced \cite{NSI1} 
as ``New Physics'' that can be searched for in neutrino oscillations. Indeed,
in matter, the first generation fermion current can be replaced
by the fermion number density in the medium: $(\overline{f}\g^\a P_X f ) \to\d^{\a0} n_f/2 $.   At finite density,  NSI therefore contribute an  effective mass  to the oscillation Hamiltonian of neutrinos:
$$
 \frac{[\Delta m^2]^{\r\s}}{E}  \sim  \sqrt{2} G_F \varepsilon_f^{\r\s} n_f~~~.
$$
 Charged current NSI, involving  a $\nu$, a charged lepton and differently charged external fermions, are also studied because they  affect the production and detection of neutrinos. However, they are not considered in this manuscript,
   where ``NSI''  is taken to mean neutral current NSI.

  The phenomenology of NSI has been widely studied (for a review, see {\it eg}
  \cite{revFT}), because they can contribute in
  neutral current neutrino scattering \cite{BR,DP-GRS,Biggio:2009nt},
  and via the matter effect to neutrino oscillations in
 Long Baseline experiments \cite{LBNL},
  the sun and  the atmosphere \cite{atm,GGetal18},
   supernovae \cite{SN1}, neutron stars \cite{NS},  and  the early
  Universe \cite{deSalasSergio,Serpico}. 
  In particular, the effects of NSI in terrestrial
  neutrino oscillation experiments have been carefully studied,
  in order  to explore the prospects of disentangling NSI
  from the  minimal set of mixing angles, masses and phases
  \cite{confusion,LBNL}.

  More recently, ``Generalised Neutrino  Interactions''(GNI)  have
  been discussed \cite{ASdRR,ATZ,FGAX,Bischer:2019ttk},
  which involve two light neutrinos,
  and two first generation fermions.  Since the neutrinos are only required to be light, but not members of an SM doublet,  GNI include scalar
  and tensor four-fermion operators involving sterile ``right-handed''
  neutrinos:
  $$
(\overline{\nu_R} _\r  \nu_{L\s})
  (\overline{f} P_X f )
  ~~ , ~~~
(\overline{\nu_R} _\r  \sigma^{\a\b}   \nu_{L\s})
  (\overline{f} \sigma_{\a\b}   P_L f )
~~~,
  $$
  where $\s^{\a\b} = \frac{i}{2}[\g^\a,\g^b]$.
  Such  scalar (and tensor) interactions are interesting, because
  the COHERENT experiment \cite{COHERENT} measured 
  neutrino scattering  on nuclei
  at  momentum transfer $\sim 30-70$ MeV,
  where the cross-section is coherently enhanced $\propto A^2$
  (where $A=$ atomic number).
  Unlike the ``matter effect'', which is a
   forward scattering amplitude so  only a vector current of
  SM neutrinos  can contribute,
  the COHERENT  cross-section  is 
  sensitive to the scalar interaction (which  is coherently enhanced), 
  as well as having reduced sensitivity to the
  tensor interaction\footnote{The
    literature contains various statements about
    coherent components of the tensor. Reference \cite{BGN},
at zero-momentum-transfer, showed that
the tensor in a polarised target can flip
    the  helicity of relativistic Dirac neutrinos, without the $m_\nu/E$
    suppression factor arising with the axial vector.
   This is not enhanced $\propto A^2$. 
    However, in
    the non-relativistic expansion of the nucleon
    current \cite{Cirelli}, there is a  coherently enhanced
    piece, suppressed by momentum-transfer.  It was
    discussed for $\mec$ in \cite{CDK}.
  }. In this manuscript,
  we focus on NSI.

  The bounds  on NSI  from
  neutrino scattering experiments \cite{BR,DP-GRS}, are 
 of order $|\varepsilon_f^{\r\s}| \lsim 0.1 \to 1$.
  A recent combined fit \cite{GGetal18}  to current oscillation data
  and the results of the COHERENT experiment  gives
  bounds $|\varepsilon_f^{\r\s}| \lsim 0.01$, except on
  the diagonal, where NSI large enough to flip the sign
  of the SM contribution are allowed\footnote{
    Oscillations
  are sensitive to the sign of the matter
  contribution, but only for  flavour differences}.
  The authors of this study assume that the flavour
  structure of NSI  on $e$s, $u$s or $d$s is the same
  (so $\varepsilon_f^{\r\s} = \varepsilon_f \varepsilon^{\r\s}$),
  and that NSI are a small perturbation around the
  standard  parameters that give best fit solutions
  in the absence of NSI. With these assumptions,
  they set constraints on NSI, meaning that
  larger values are excluded.
  The results of the
  COHERENT experiment 
  are an important input to this analysis, because
  the oscillation data is sensitive to differences in the eigenvalues
  of the  propagation Hamiltonian, whereas
the COHERENT results constrain  the neutral current scattering rate.
So large flavour-diagonal NSI are constrained by COHERENT. 
  The COHERENT constraints alone, without
assumptions about the flavour structure of $\varepsilon$,
are discussed in \cite{Giunti:2019xpr}.

  In the Standard Model, neutrinos share an SU(2) doublet
  with charged leptons, so that SM gauge-invariant
  operators that mediate   NSI may
  also mediate  stringently constrained,
  charged lepton flavour changing processes.
  For instance, the contact   interaction 
  of eqn (\ref{eqn1}), for $f=e_L$, could be generated by
  the dimension six operator
\beq
  -2\sqrt{2} G_F \varepsilon_e^{\r\s}
  (\overline{\ell}_\r\g^\a\ell_\s)  (\overline{\ell}_e\g_\a\ell_e) 
\label{NSIopex}
  \eeq
where $\bar{\ell}$ is the SU(2) doublet $(\overline{\nu_L}, \overline{e_L})$. However,
 this operator also induces the four-charged-lepton interaction
$(\overline{e}_\r\g^\a P_L  e_\s)  (\overline{e }\g_\a P_L e) $
whose coefficient would be strictly constrained by
decays $e_\s \to e_\r e\bar{e}$.  These concerns
can be avoided by instead constructing NSI at dimension
eight in the SMEFT
, for instance as
  \beq
- 2\sqrt{2} G_F  \varepsilon_f^{\r\s}(\overline{\nu} _\r \g_\a  \nu_\s)
(\overline{f}\g^\a f )~~ \longleftarrow~~
\frac{
C_f^{\r\s}}{\Lambda_{NP}^4}
(\overline{\ell}^p_\r \epsilon_{pQ}H^{Q*}) \g_\a (H^R \epsilon_{Rs}\ell^s_\s)(
\overline{f}\g^\a f )
  \label{intro3}
  \eeq
  where $\epsilon_{pQ}$ is the antisymmetric SU(2) contraction
  given in eqn (\ref{eqnApp1}).
  When the Higgs $H= (H^+,H_0)$ takes a vacuum expectation
  value $\langle H_0 \rangle = v$,  the dimension eight operator reproduces
  the contact interaction on the left (this is discussed in more detail
  in section \ref{sec:notn}), with
\beq
\varepsilon_f^{\r\s} = C_f^{\r\s} \frac{v^4}{\Lambda_{NP}^4}~~.
\label{epsI}
\eeq
It is clear that to obtain $\varepsilon\gsim 10^{-3}$,   the
New Physics scale $\Lambda_{NP}$ cannot be  far above the weak scale and is likely to be within the reach of the LHC.

Models that generate such large effects in the neutrino
  sector, while avoiding the stringent bounds on
  charged Lepton Flavour Violation(LFV) \cite{PSI}, have been explored by
  various authors\footnote{Reference  \cite{BabuEtal} is a recent study of tree-NSI models that are not engineered to avoid tree-level LFV.}.
  The authors of  \cite{GXOW} 
  considered the case where  NSI were generated
  at tree level by the exchange of
  new particles of mass $\gsim m_W$, and required that
  the heavy mediators not induce  tree-level
  LFV  interactions  at dimension
  six or eight. They allowed for cancellations among
the   mediators of  operators of a given dimension, but
not for cancellations between the coefficients of
operators of different dimension, and  found various
viable models. Similarly, reference \cite{Antusch}
 considered models with heavy new particles that
induced NSI at tree level, however these authors  did not allow cancellations
among the contributions of different mediators to LFV
interactions. They showed that their allowed  models induced
additional, better constrained operators, so that
$\varepsilon \gsim 10^{-2}$ was excluded.
In this manuscript,  we  review this
question  from an EFT
perspective  allowing  arbitrary cancellations,
also between operators of dimension six and eight\footnote{
  Cancellations between operators of
  different dimension occur already in the SM:
the Higgs potential minimisation relates the dimension
two  operator $-M^2 H^\dagger H$ to $\lambda (H^\dagger H)^2/2$.},
in order to find linear combinations of operators that
induce NSI but not LFV at tree level.

Models with light mediators  have also  been
constructed \cite{PP,FarzanModel,FarzanModel2}. Such models
are motivated, because  a detectable $\varepsilon$
cannot be  small, suggesting that
 any heavy mediator   could be  within the
range of the LHC. 
The models of \cite{PP,FarzanModel} involve a
 light ($\gsim 10 $ MeV)  feebly coupled $Z'$,
 which can avoid tree-level LFV constraints
 by a suitable choice of couplings; in \cite{FarzanModel2},
 the SM neutrinos share mass terms with additional
 singlets, which are charged under the U'(1).

  Even if the New Physics responsible for NSI does not
  induce LFV at tree level, loop effects
  could  mix  NSI and  LFV operators. 
  Reference \cite{DP-GRS} considered a particular
  dimension eight NSI operator, and  erroneously argued that the exchange
of  a $W$ boson between the two neutrino legs
would   transform them into charged leptons, 
thereby  inducing  a contact interaction that was
  severely constrained by experimental bounds on
  charged Lepton Flavour Violation (LFV). 
  However, it was pointed out  in \cite{Biggio}, that the 
  log-enhanced, one-loop  mixing  of this  NSI operator
  into LFV operators
  vanished. The apparent conclusion
  was that at one loop,   there is no  model-independent constraint
  on NSI from LFV.

  In this manuscript, we  revisit the 
  EFT description of NSI, and the LFV it induces via
  electroweak loops. 
 We are therefore  neglecting models with
 light mediators, and our results apply 
 when NSI are present as a contact interaction
 above the weak scale, where the usual SMEFT can
 be applied.  
 In  section \ref{sec:notn}, we introduce the 
 two sets of operators    that we will use in the analysis:
 SU(2)-invariant operators for the EFT  above the weak scale,
 and  QED$\times$QCD invariant operators below $m_W$.
 Also, the matching between the bases
 is given and the  operator combinations that induce
 either NSI, or LFV, at low energy  are listed.
 Section  \ref{sec:loops} is about
 Renormalisation Group Equations (RGEs),
 which encode 
the Higgs and $W$ loops that mix  NSI  and LFV  
 at scales  above $m_W$. 
 In this manuscript, we limit ourselves to one-loop
 RGEs\footnote{Recall that
 the loop corrections obtained
 with one-loop RGEs  occur in all
 heavy-mediator models, and are
 independent of the renormalisation
 scheme used for the operators that
 are introduced to mimic the interactions
 induced by high-scale particles.}, which describe the log$^n$-enhanced  part of
 all $n$-loop diagrams.
 The one-loop  RGEs  are known for dimension six  operators \cite{JMT}, and
 those for  our dimension eight operators  
 are obtained in section \ref{sec:loops}.
 Finally,
  in  the results section \ref{sec:results},
  which should be accessible without reading
  the more technical section \ref{sec:loops},
  we   show that in  most cases,
  the operator combinations that at tree
  level match onto NSI without LFV, induce LFV at one loop
  via the RGEs. The resulting sensitivities of  LFV processes
  to NSI  are given.
   We summarise
 in section \ref{sum}.


\section{ Operators }
\label{sec:notn}

\subsection{In the $SU(3)\times SU(2)\times U(1)$ theory above $m_W$}
\label{ssec:2.1}

We  suppose  a New Physics model at a scale
$\Lambda_{NP} > m_W$, that
induces  lepton-flavour-changing vector   operators
 of  dimension  six and eight,  which  at tree level  generate  (neutral current) NSI
 but no LFV.
  We want  to know whether Higgs or $W$ loops could mix
 such  operators into LFV operators, so
 we need a list of NSI/LFV  vector operators of dimension eight and six.
 These operators will be added to the SM Lagrangian
 as ${\cal L}_{SM} \to
 {\cal L}_{SM} + \delta {\cal L}$, with 
 \bea
 \delta {\cal L}& = & \sum_{O,\zeta} \frac{C_O^\zeta}{\Lambda_{NP}^{2n}}{\cal O}_O^\zeta + h.c.
\eea
where $n= 1$ or 2  for respectively  dimension six or eight  operators,
 $\{O\}$  is the basis of  operators with Lorentz structure $\g_\a
\otimes \g^\a$,
and  $\zeta$ represents the flavour indices $\r\s ff$.
To avoid  cluttering the notation, the flavour indices are sometimes
reduced to $\r\s$ or suppressed. 
Greek indices from the
beginning of the alphabet ($\a ,\b $...) are Lorentz indices,
and those from the
end of the alphabet ($\s ,\r $...) are charged lepton flavour indices.
The New Physics scale $\Lambda_{NP}$ is required to be above $m_W$,
but is otherwise undetermined, being one of the parameters
controlling the size of $\varepsilon_f^{\r\s}$ (see eqn \ref{epsI}).
In later sections,  loop effects containing $\ln (\Lambda_{NP}/m_W)$
will arise, which we conservatively take $\simeq 1$.

The  Higgs doublet is written
\begin{equation}
{H} = \left(
\begin{array}{c}
H^+\\
H_0
\end{array}
\right) \to  \left(
\begin{array}{c}
0\\
v
\end{array}
\right) 
\end{equation}
where after the arrow is the vacuum expectation value
 with $1/v^2 = 2\sqrt{2} G_F$, and the Higgs is included
in the Standard Model Lagrangian (in the mass eigenstates of charged leptons) as
\bea
 {\cal L}_{SM} &=&
 \overline{\ell} i \Dslash \ell +... 
 -\{y^\rho_e \overline{\ell}_\rho H e^\rho_{R}  + h.c.\} 
+ (D_\mu H)^\dagger D^\mu H  - M^2H^\dagger H +\frac{\lambda}{2} (H^\dagger H)^2~~~.
\label{LSM}
\eea
where the physical Higgs mass $\simeq 125$ GeV is $m^2_h = \lambda v^2$, which
corresponds to $\lambda\simeq 1/2$.
At tree level, the minimum of the Higgs potential   is given by
\beq
M^2 - \lambda v^2 = 0 ~~,
\label{mincondtree}
\eeq
and the one-loop minimisation is discussed in Appendix \ref{app:matching}.
Since we  will write
RGEs for operators of dimension six and eight,  which can mix due
to Higgs mass insertions,
we will frequently use a parameter
\beq
\eta \equiv \frac{M^2}{\Lambda_{NP}^2}~~ , ~~ \frac{\eta}{\lambda} =
 \frac{v^2}{\Lambda_{NP}^2}~.
\label{eta}
\eeq

Consider first to construct  operators 
involving doublet leptons  and
 SU(2) singlet external fermions $f$.
The dimension six vector operator of the ``Warsaw'' basis \cite{polonais}
is
\bea
 {\cal O}^{ \r\s}_{M2,f}  \equiv
(\overline{\ell}_\r  \g_\a   \ell_\s)(\overline{f}  \g^\a  f) ~~~,
\label{opsa}
\eea
 referred
to as ``M2'', because the dimension eight operators will mix into
it via  insertions of the Higgs mass parameter $M^2$.
At dimension eight, 
a convenient basis is
 \bea
{\cal O}^{ \r\s}_{NSI,f} \equiv (\overline{\ell}_\r \epsilon H^{*}) \g_\a
(H \epsilon \ell_\s)(\overline{f}  \g^\a  f) ~~~,~~~
{\cal O}^{ \r\s}_{H2,f}  \equiv
 (\overline{\ell}_\r H \g_\a  H^\dagger  \ell_\s)(\overline{f}  \g^\a  f)~~~,
\label{opsb}
\eea
where $\epsilon$ is the totally anti-symmetric tensor in two dimensions.
There could be
additional operators with derivatives, but we neglect the
Yukawa couplings, in which limit the derivative operators vanish by
the equations of motion. 


For the case where the external fermions are
SU(2) doublets, the Warsaw basis (of  dimension six operators)
contains 
${\cal O}_{M2,f}$ for $f\in \{\ell,q\}$, and also the
triplet contraction $(\overline{\ell}_\r\vec{\tau}  \g_\a  \ell_\s)
(\overline{q} \vec{\tau} \g^\a  q)$. The analogous four-lepton
triplet contraction is not included, because it can
be rewritten:
$$
(\overline{\ell}_\mu\vec{\tau}  \g_\a  \ell_\tau)
(\overline{\ell}_e \vec{\tau} \g^\a  \ell_e) =
2(\overline{\ell}_\mu  \g_\a  \ell_e)
(\overline{\ell}_e  \g^\a  \ell_\tau)
-(\overline{\ell}_\mu  \g_\a  \ell_\tau)
(\overline{\ell}_e  \g^\a  \ell_e)~~.
$$
The singlet operators are more   convenient for matching to low-energy
four-fermion operators than the triplets, so we
make a similar transformation for
the  triplet operator involving quarks, and take at
 dimension six  for external doublet quarks:
\bea
 {\cal O}^{ \r\s}_{M2,q}  \equiv
(\overline{\ell}_\r  \g_\a   \ell_\s)(\overline{q}  \g^\a  q)
~~~,~~
{\cal O}^{ \r\s}_{LQM2,q}  \equiv
(\overline{\ell}_\r  \g_\a   q)(\overline{q}  \g^\a  \ell_\s) ~~~.
\label{opsc}
\eea

At dimension eight, 
Rossi and Berezhiani \cite{BR}  propose 
five  operators
\bea
{\cal O}^{\r\s}_S= (\overline{\ell}_\r  \g_\a  \ell_\s)(\overline{q}  \g^\a  q)
 (H^\dagger  H) &&
{\cal O}_{TLH}^{\r\s} = (\overline{\ell}_\r \tau^a \g_\a  \ell_\s)(\overline{q} \g^\a  q)
 (H^\dagger \tau^a H) \nonumber\\
  {\cal O}_{TQH}^{\r\s} = (\overline{\ell}_\r  \g_\a  \ell_\s)(\overline{q}  \g^\a \tau^a  q)
 (H^\dagger\tau^a  H) &&
 {\cal O}_{TLQ}^{\r\s} = (\overline{\ell}_\r \tau^a \g_\a  \ell_\s)(\overline{q} \tau^a \g^\a  q)
 (H^\dagger  H) \nonumber\\
 (\overline{\ell}_\r \tau^a \g_\a  \ell_\s)(\overline{q} \tau^b \g^\a  q)
 (H^\dagger \tau^c H) \epsilon_{abc} &\equiv &  {\cal O}^{\rho\s}_{TTT}
\label{BRdim8}
\eea
where to be concrete, the  external
fermion is taken to be a first generation
quark doublet.  The first two operators would be present
for singlet external currents.

In order to count 
the number of operators, notice
that it corresponds to
the number of independent SU(2) contractions for
an operator constructed from the fields:
$$
 (\overline{\ell}^i_\r   \g_\a   \ell^j_\s)(\overline{q}^k  \g^\a q^l )  (H^{\dagger M}    H^N)
$$
where $\{i,j,k,l,M,N\}$ are SU(2) indices.
The possible contractions involve three $\tau$s,
one $\delta$ and two $ \tau$s,
one $\delta$ and two $ \epsilon$s,
or three $\delta$s. But the $\tau \tau\tau$,  $\delta \tau\tau$
and $\delta \epsilon\epsilon$ contractions can be rewritten
as three $\delta$s using  the  Fierz or SU(2) identities
given in eqn (\ref{Fierz}).
Then there are six  $\d\d\d$ contractions,
among which we find one relation, leaving
five independent operators
(This is discussed in more detail in
Appendix \ref{app:basis}).

It is  convenient  to use an alternative basis without
triplet contractions $ {\cal O}^{\rho\s}_{TTT}$, to simplify the matching onto  the
Higgsless theory below $m_W$. The dimension
six operators  in our basis,
  in the case where the external fermion
  is the first generation quark doublet $q$, are
 \bea
{\cal O}^{ \r\s}_{M2,q} \equiv
 (\overline{\ell}_\r \g_\a   \ell_\s) (\overline{q}  \g_\a  q)
 ~~~&,&~~~
{\cal O}^{ \r\s}_{LQM2,q} \equiv
 (\overline{\ell}_\r \g_\a   q) (\overline{q}  \g_\a  \ell_\s)
\label{opsd6us}
\eea
where the SU(2) contractions
are inside parentheses.
At dimension eight, we take
\bea
{\cal O}^{ \r\s}_{NSI,q} \equiv (\overline{\ell}_\r \epsilon H^{*}) \g_\a
(H \epsilon \ell_\s)(\overline{q}  \g^\a  q) ~~~&,&~~~
{\cal O}^{ \r\s}_{H2,q}  \equiv (\overline{\ell}_\r H) \g_\a  (H^\dagger  \ell_\s)
(\overline{q}  \g^\a  q) \nonumber\\
{\cal O}^{ \r\s}_{CCLFV,q} \equiv
 (\overline{\ell}_\r \g_\a   q) (\overline{q} H) \g_\a  (H^\dagger \ell_\s) 
 ~~~&,&~~~
[{\cal O}^\dagger_{CCLFV,q}] ^{ \r\s} \equiv
 (\overline{\ell}_\r H) \g_\a (H^\dagger  q) (\overline{q} \g_\a   \ell_\s) 
\label{opsSU2}\\
{\cal O}^{ \r\s}_{CCNSI+,q} \equiv ({\cal O}^{ \r\s}_{CCNSI,q} +
[{\cal O}^{\dagger}_{CCNSI,q}]^{\r\s})
&\equiv&
 (\overline{\ell}_\r \g_\a   q) (\overline{q} \epsilon H^{*} ) \g_\a
 (H \epsilon \ell_\s)
+ 
(\overline{\ell}_\r \epsilon H^{*})  \g_\a  (H \epsilon  q)  (\overline{q}  \g_\a
 \ell_\s) 
\nonumber
\eea
where the  SU(2) contractions are
inside the parentheses.  The relation of this
basis to the Berezhiani-Rossi basis
is discussed in appendix \ref{app:basis}. 

The operators ${\cal O}_{H2}$ and ${\cal O}_{NSI}$ 
are hermitian(as matrices in lepton  flavour space), as is the combination 
${\cal O}_{CCNSI} + {\cal O}_{CCNSI}^\dagger$ (which
corresponds to one of the $\d\d\d$ contractions
discussed above).
The remaining two operators,
${\cal O}_{CCLFV}$ and ${\cal O}_{CCLFV}^\dagger$,
are not hermitian, but appear in the one-loop
RGEs in the  combination
${\cal O}_{CCLFV,+} \equiv {\cal O}_{CCLFV} + {\cal O}_{CCLFV}^\dagger$. As a result, our
basis of dimension eight operators for external doublets  contains
only four operators  that mix with each other.
An additional operator, ${\cal O}_{CCLFV} - {\cal O}_{CCLFV}^\dagger$, 
decouples from the operator mixing but is included in
our basis for completeness.
The  matching of these operators onto low energy operators
is given in table
\ref{tab:ops}.

\begin{table}[h]
\centering
\begin{tabular}{l|l|l|l}
name  && operator &  below $m_W$\\
\hline
${\cal O}^{ \r\s}_{NSI,q} $&x&$ (\overline{\ell}_\r \epsilon H^{*}) \g_\a
(H \epsilon \ell_\s)(\overline{q}  \g^\a  q) $&
$- v^2 (\overline{\nu}_\r  \g_\a P_L  \nu_\s)(\overline{q}  \g^\a  q) $
\\
${\cal O}^{ \r\s}_{H2,q}  $&x&$ (\overline{\ell}_\r H) \g_\a  (H^\dagger  \ell_\s)
(\overline{q}  \g^\a  q)$&
$ v^2 (\overline{e}_\r \g_\a  P_L  e_\s)
(\overline{q}  \g^\a  q)$
\\
${\cal O}^{ \r\s}_{M2,q} $&x&$
 (\overline{\ell}_\r \g_\a   \ell_\s ) (\overline{q}  \g^\a  q)
 $&$
 (\overline{e}_\r \g_\a P_L  e_\s+\overline{\nu}_\r \g_\a   \nu_\s ) (\overline{q}  \g^\a  q) $ \\
\hline
 \hline
${\cal O}^{ \r\s}_{LQM2,q} $&&$
 (\overline{\ell}_\r \g_\a   q) (\overline{q}  \g^\a  \ell_\s)
 $&
 $
 (\overline{\nu}_\r \g^\a  P_L\nu_\s)(\overline{u} \g_\a P_L  u )
+ (\overline{e}_\r \g^\a P_L  e_\s )(\overline{d} \g_\a  P_L d )
$ \\
 &&& $
+ (\overline{\nu}_\r\g^\a P_L  e_\s )( \overline{d} \g_\a P_L  u )
+(\overline{e}_\r  \g^\a  P_L\nu_\s ) (\overline{u} \g_\a  P_L d)
$ \\
 \hline
${\cal O}^{ \r\s}_{CCLFV,q} $&&$
  (\overline{q} H) \g_\a  (H^\dagger \ell_\s) (\overline{\ell}_\r \g^\a   q)
$&$
  2 v^2( \overline{e}_\r \g_\a  P_L e_\s)  (\overline{d}  \g^\a P_L  d )
$ \\
$~+[{\cal O}^\dagger_{CCLFV,q}] ^{ \r\s} $&&$
 ~+ (\overline{\ell}_\r H) \g_\a (H^\dagger  q) (\overline{q} \g^\a   \ell_\s) 
$& $
+  v^2  (\overline{e}_\r\g_\a  P_L \nu_\s ) (\overline{u} \g^\a P_L d)  
 +  v^2 (\overline{\nu}_\r \g_\a  P_L e_\s)  (\overline{d}  \g^\a  P_L u)
 $ \\
 \hline
${\cal O}^{ \r\s}_{CCNSI,q} $&x&$
 (\overline{q} \epsilon H^{*} ) \g_\a
 (H \epsilon \ell_\s)  (\overline{\ell}_\r \g^\a   q)
 $& $
  - 2v^2(\overline{\nu}_\r    \g_\a  P_L  \nu_\s)  (\overline{u}  \g^\a 
P_L u )
$ \\
$~+ [{\cal O}^{\dagger}_{CCNSI,q}]^{\r\s} $& &$
 ~+ (\overline{\ell}_\r \epsilon H^{*})  \g_\a  (H \epsilon  q)  (\overline{q}  \g^\a
 \ell_\s)
$& $
 -
 v^2   (\overline{e}_\r \g_\a  P_L  \nu_\s) (\overline{u} \g^\a P_L d)
 -  v^2(\overline{\nu}_\r    \g_\a  P_L   e_\s) ( \overline{d}  \g^\a  P_L u)
$ \\
 \hline
${\cal O}^{ \r\s}_{CCLFV,q} $&&$
  (\overline{q} H) \g_\a  (H^\dagger \ell_\s) (\overline{\ell}_\r \g^\a   q)
$&$
   v^2 (\overline{\nu}_\r \g_\a  P_L e_\s)  (\overline{d}  \g^\a  P_L u)
$ \\
$~-[{\cal O}^\dagger_{CCLFV,q}]^{ \r\s} $&&$
 ~- (\overline{\ell}_\r H) \g_\a (H^\dagger  q) (\overline{q} \g^\a   \ell_\s) 
$& $
-  v^2  (\overline{e}_\r\g_\a  P_L \nu_\s ) (\overline{u} \g^\a P_L d)  
  
 $ \\
\end{tabular}
\caption{SMEFT operators used  in the RGEs of  this paper,
and four-fermion operator below $m_W$ onto which
they match. For concreteness,   the
external fermion $f$ is taken to be  a quark doublet $q$.
The first three operators are present for
all external fermions; those below the double line
are only required  for external doublets
when they are quarks, or   leptons
with  $(\r,\s) \in \{(\tau, \mu), (\mu,\tau)\}$.
For external doublet leptons ($q\to \ell_e$ in
  the table), when
$ \r=e$ or $\s=e$, 
only the  operators with a cross in the second column
are required, and notice that below $m_W$
($u\to \nu_e$ and  $d\to e$ in
  the table),
${\cal O}_{CCNSI+,\ell_e}$ matches onto a $4\nu$ operator,
an NSI operator and a CC operator  after a Fierz transformation. 
\label{tab:ops}}
\end{table}

Finally, if the external doublets are  leptons $\ell_e$,
the flavour indices of the
operators  can be  $\{\r,\s\}  \in \{\mu,\tau\}$,
or  one of $\r,\s$ can be $e$.
In the case  $\{\r,\s\}   = \{\mu,\tau\}$, 
there are no identical fermions, and the
 basis given above for doublet quarks can be used.

For the case where one of  $\r,\s$ is $e$, there are some redundancies. First,
notice
that  in this case, the operator only carries one flavour index, which
can be  taken to be $\s \in \{\mu, \tau\}$.
Then inequivalent operators that annihilate $\ell_\s$ can be constructed,
and the $+h.c.$ will look after the operators which create $\ell_\s$ .
One finds the following equalities:
\bea
{\cal O}^{ e\s}_{CCNSI,\ell}
= {\cal O}^{ e\s}_{NSI,\ell}
~~~,~~
{\cal O}^{ e\s}_{CCLFV,\ell}
= {\cal O}^{ e\s}_{H2,\ell}
~~~,~~
{\cal O}^{ e\s}_{LQM2,\ell}
 =
{\cal O}^{ e\s}_{M2,\ell}
\label{redundancies}
\eea
and the relation
\bea
 {\cal O}^{ e\s}_{CCNSI,\ell}-[{\cal O}^{\dagger }_{CCNSI,\ell}]^{ e\s}  & =
& {\cal O}^{ e\s}_{CCLFV,\ell}
-[{\cal O}^{\dagger }_{CCLFV,\ell}]^{ e\s}
\eea
so that a sufficient basis in this case should be
\bea
{\cal O}^{ e\s}_{M2,\ell} &\equiv&
 (\overline{\ell}_e \g_\a   \ell_\s) (\overline{\ell}_e  \g_\a  \ell_e)
 \nonumber\\
{\cal O}^{ e\s}_{NSI,\ell}& \equiv &(\overline{\ell}_e \epsilon H^{*}) \g_\a
(H \epsilon \ell_\s)(\overline{\ell}_e  \g^\a  \ell_e)
\nonumber\\
{\cal O}^{ e\s}_{H2,\ell}& \equiv &(\overline{\ell}_e H) \g_\a  (H^\dagger  \ell_\s)
(\overline{\ell}_e  \g^\a  \ell_e)\nonumber\\
{\cal O}^{ e\s}_{CCNSI+,\ell}
& \equiv&
(\overline{\ell}_e \g_\a   \ell_\s) (\overline{\ell_e} \epsilon H^*) \g_\a
(H \epsilon \ell_e) +  (\overline{\ell}_e \g_\a   \ell_e) (\overline{\ell_e} \epsilon H^*) \g_\a  (H \epsilon \ell_\s)
\label{ellbasis}
\eea
with $\s$ ranging over $\{ \mu,\tau\}$.

\subsection{In the $QCD \times QED$ theory below $m_W$}
\label{ssec:<mW}

At $m_W$, 
the  $SU(3) \times SU(2) \times U(1)$-invariant SMEFT
is matched onto an effective theory that  is QCD$\times$QED invariant,
where NSI operators can no longer  mix to LFV operators.
The dimension six and eight SMEFT operators  all match onto four fermion
operators of the low energy theory,
which, for LFV (and Charged Current) operators,
are defined   with Lorentz structure and chirality subscripts,
and flavour superscripts:
\beq
    {\cal O}_{V,XY}^{\tau\mu ff}   = (\overline{\tau} \g^\mu P_X \mu)
     (\overline{f} \g_\mu P_Yf)~~~,
\label{belowmW}
    \eeq
and are added to the Lagrangian as $\delta {\cal L}=
2\sqrt{2} G_F C_{V,XY}^{\r\s\a\b} {\cal O}_{V,XY}^{\r\s\a\b}$.
However,  the low energy  NSI  coefficients
are defined with opposite sign
to agree with the 
convention that  NSI operators have the same sign  
as the Fermi interaction (see eqn \ref{eqn1}).

The third column of table \ref{tab:ops}  gives the  combination
of low-energy operators onto which a given SMEFT operator
is matched at tree level. This table shows that
for  external fermions other than the  quark doublet,
there is  at low energy
only one LFV operator, and one NSI operator
(for an external quark doublet, there are  two
of both,  involving $u_L$ and $d_L$) in the theory below $m_W$.
The  coefficients of the low-energy operators will
be a sum of SMEFT coefficients, so for
a given external fermion $f\in\{e_L,e_R, u_L,u_R, d_L,d_R\}$
there is only one combination of SMEFT coefficients
that needs to  be non-zero, and another than should vanish, in order to have NSI
without LFV at tree level. In the remainder of this
subsection,  for each possible  external fermion,
we give these combinations of SMEFT coefficients.

Three comments about these directions in coefficient space:
first, in the low energy theory, we allow
tree-level charged current operators, in the perspective
that the bounds on flavour-changing charged current processes
are  not more restrictive than the $\varepsilon \lsim 0.01$
bounds on NSI \cite{GGetal18}. 

Secondly, 
arbitrary cancellations among operators
of same  and different
dimension are allowed. This  differs from the
studies of, {\it eg}, References \cite{GXOW,Antusch},
who constructed New Physics models to generate
the SMEFT operators,  then restricted to the
cancellations that the authors  considered natural.
In the EFT perspective of this manuscript,
cancellations among operators
of the same dimension are allowed because
they just reflect the choice of operator basis.
Cancellations among  four-fermion operators
of dimension six and eight are also allowed
 because a similar cancellation
between operators of different dimension
occurs in minimising  the
Higgs potential(see eqn \ref{mincondtree}).
Cancellations between contributions of different
  power
of $\log (\Lambda_{NP}/m_W)$ are however
not allowed (this is further discussed
in section \ref{ssec:cancellations}).

Thirdly, the results  listed here are well-known; the purpose of
this discussion  is to give the conditions in the
 operator basis used here.  For instance, 
low-energy  LFV cancels  between
$C^{\mu\tau}_{M2,\ell}$ and $C^{\mu\tau}_{LQM2,\ell}$
if $C^{\mu\tau}_{M2,\ell} = -C^{\mu\tau}_{LQM2,\ell}$.
This  could be written as
$$
\varepsilon^{\mu\tau}_{3,\ell\ell} = -\varepsilon^{\mu\tau}_{\ell\ell}
$$
in  a  basis\footnote{
  Although the ``triplet''  4$\ell$ operator is absent
  from the Warsaw basis, it is not redundant in
  a basis where the
first generation indices are required to be  in the
second operator current.}
  which included
${\cal O}_{3,\ell\ell}^{\mu\tau}$ = $(\overline{\ell}_\mu \vec{\tau} \g_\a \ell_\tau)
  (\overline{\ell}_e \vec{\tau} \g^\a \ell_e)$ and
  ${\cal O}_{\ell\ell}^{\mu\tau}$ = $(\overline{\ell}_\mu  \g_\a \ell_\tau)  (\overline{\ell}_e  \g^\a
  \ell_e)$. This cancellation reflects the model-building possibility
  of putting an $L=2$ scalar dilepton $D$, with vertices
  $y_{\mu e} \overline{\ell^c}_\mu\epsilon \ell_e D$ and  
  $y_{\tau e} \overline{\ell^c}_\tau\epsilon \ell_e D$, which generates
  the contact interaction
$(\overline{\ell}^i_\mu  \g_\a \ell^k_\tau)  (\overline{\ell}^j_e  \g_\a
  \ell^l_e)\epsilon_{ij}\epsilon_{kl}$
transformable to either of the  cancelling
  combination of operators by using the  identities of eqn
  (\ref{Fierz}).

  In the case of operators with singlet external fermions,
 ${\cal O}_{NSI,f}$ induces only NSI, 
  ${\cal O}_{H2,f}$ only  LFV, and
  and  ${\cal O}_{M2,f}$ induces both.
 The  tree-level  LFV
 and NSI coefficients can be read  from table \ref{tab:ops}:
 \bea
C_{V,LR}^{\r\s ff} =
\frac{v^2}{\Lambda^2} \left( C^{\r\s}_{M2}  +C^{\r\s}_{H2} \frac{\eta}{\lambda} \right)  &~~,~~&
 \varepsilon^{\r\s}_f =
\frac{v^2}{\Lambda^2} \left(
- C^{\r\s}_{M2} + C^{\r\s}_{NSI}\frac{\eta}{\lambda}\right) 
\label{match}
\eea
where we used the tree-level Higgs  minimisation condition
${v^2}/{\Lambda^2} = \eta/\lambda$. 
So  low energy  LFV  vanishes at tree level if
  \bea
 \eta C_{H2} + \lambda C_{M2} = 0 ~~~.
\label{vanishf}
\eea
 A third interesting  coefficient combination, independent
 of those that induce NSI and LFV, is
 $ \eta C_{H2} =-\eta C_{NSI }=  -\lambda C_{M2}$, which
  induces no  low-energy interactions.

  For   external fermions that are  doublet quarks,
  NSI are proportional to
  \bea
\varepsilon^{\r\s}_{d_L}&=& \frac{v^2}{\Lambda^2}\left(-C^{\r\s}_{M2,q}  +\frac{\eta}{\lambda}
        C^{\r\s}_{NSI,q } \right)  \nonumber\\
  \varepsilon^{\r\s}_{u_L}&=&
  \varepsilon^{\r\s}_{d_L}+  \frac{v^2}{\Lambda^2}\left(-C^{\r\s}_{LQM2,q}
  +2\frac{\eta}{\lambda}
       C^{\r\s}_{CCNSI+,q }\right)
     ~~~.\label{NSIq}
  \eea
Low-energy  LFV  is induced  on $u_L$ currents by 
 ${\cal O}_{H2,q}$  and 
and ${\cal O}_{M2,q}$, and on  and $d_L$ currents by 
${\cal O}_{CCLFV+,q}$,
${\cal O}_{H2,q}$,  ${\cal O}_{M2,q}$ and ${\cal O}_{LQM2,q}$,
so the LFV coefficients are 
\bea
C_{V,LL}^{\r\s uu} &= &
\frac{v^2}{\Lambda^2} \left( \frac{\eta}{\lambda } C^{\r\s}_{H2,q}  +C^{\r\s}_{M2,q}\right)
\nonumber   \\
C_{V,LL}^{\r\s dd} &= & C_{V,LL}^{\r\s uu}+ \frac{v^2}{\Lambda^2} \left(
2\frac{\eta}{\lambda }  C^{\r\s}_{CCLFV+,q} +C^{\r\s}_{LQM2,q}\right) 
 ~~~.
\label{vanishq}
\eea
It is straightforward to check from table \ref{tab:ops}
that there are two other independent combinations, that do not
induce any low-energy operators, due to cancellations.

Finally, when the  external fermion is a doublet lepton
  and the flavour indices are $\r,\s \in \{( \tau,\mu), (\mu,\tau)\}$,
the  low energy NSI and  LFV coefficients are
\bea
\varepsilon^{\r\s}_{e_L}&=& \frac{v^2}{\Lambda^2}\left(-C^{\r\s}_{M2,\ell}  +\frac{\eta}{\lambda}
        C^{\r\s}_{NSI,\ell } \right)  \label{nsile}\\
C_{V,LL}^{\r\s ee} &= &
\frac{v^2}{\Lambda^2} \left( 
\frac{\eta}{\lambda} (  C^{\r\s}_{H2,\ell}+  2C^{\r\s}_{CCLFV+,\ell} )
+  C^{\r\s}_{LQM2,\ell}+  C^{\r\s}_{M2,\ell}\right) ~~~.
\label{vanishl}
\eea
In the case where one of  $\r,\s$ is an electron,  LFV
vanishes when the condition (\ref{vanishf}) applies, and
  \bea
\varepsilon^{e\s}_{e_L}&=& \frac{v^2}{\Lambda^2}\left(-C^{e\s}_{M2,q}  +\frac{\eta}{\lambda}
       ( C^{e\s}_{NSI,q }+
       C^{e\s}_{CCNSI+,q })\right)
     ~~~.\label{NSIlid}
  \eea


\section{Loop diagrams and the Anomalous Dimension matrices}
\label{sec:loops}

We consider the  mixing among the operators listed
in the first column of table \ref{tab:ops}, due  to the one-loop diagrams
induced by $W$ or Higgs exchange that are illustrated
in figures \ref{fig:3},\ref{fig:4}, \ref{fig:penguins}, and
\ref{fig:5}.   There are additional wavefunction
diagrams that are not illustrated. The loops
involve the SU(2) gauge coupling $g$ and
Higgs self-interaction $\lambda$;  Yukawa couplings are neglected
because they are small for leptons and first generation
fermions. The hypercharge interactions are less
interesting, because they cannot change the  SU(2)
structure of the operators. They are included,
for illustration,  for external singlet fermions. 
The calculation is performed  in  $\overline{MS}$  in $R_\xi$ gauge,
with the Feynman rules of unbroken SU(2), partially given
in appendix \ref{sec:ids}.

\subsection{Diagrams and divergences for gauge bosons}

\begin{figure}[ht]
\vspace{-.2cm} 
 \begin{center}
\epsfig{file=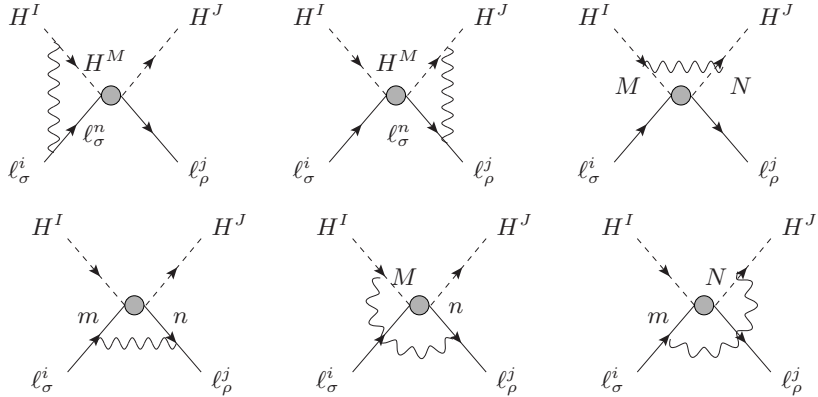,width=0.6\textwidth}
 \end{center}
\vspace{-.5cm} 
\caption{ $W$ loop corrections to
  operators represented by the grey circle;
 there is also a current of  external fermions $f$
 present in the operator, but these 
  lines are not drawn  because
  they do not participate in the loop. 
  These
  diagrams occur for all dimension eight operators;
   there
  are in addition wavefunction diagrams. 
  Only the fourth diagram (without the Higgs legs), and wavefunction
  diagrams are  present for
  dimension six  operators\label{fig:3}.
 Superscripts
are SU(2) indices, subscripts are flavour indices.
}
\end{figure}

Consider first the diagrams of figure \ref{fig:3}, which  could 
contribute  to the running and mixing of all dimension eight operators.
The fermion wavefunction diagrams are $\propto \xi$ (the parameter
of R-$\xi$ gauge), and the $W$  corrections to a scalar
leg give a divergence
\bea
(-3+\xi)\frac{ g^2}{4}[\tau^a\tau^a]_{IJ}
\frac{i}{16\pi^2 \epsilon}  p^2 ~~~.
\eea
We systematically check  that the coefficients of $\xi$
vanish in our calculation, so in the following, we drop all
the diagrams which are proportional to $\xi$. 
Indeed, all the vertex diagrams in
figure \ref{fig:3} are $\propto \xi$, so they
do not contribute. Only the divergence from the
scalar wavefunction remains, which renormalises operators
but does not mix them among each other. 

\begin{figure}[thb]
\begin{center}
\epsfig{file=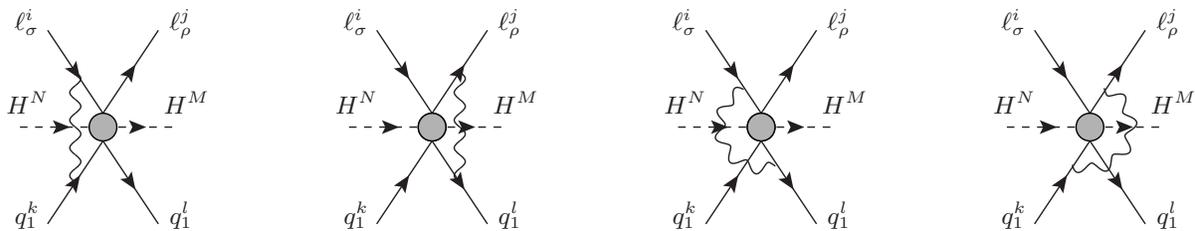,width=0.88\textwidth}
\end{center}
\vspace{-.5cm} 
\caption{$W$ loops  that can arise when the
  external fermion is an SU(2) doublet. Superscripts
are SU(2) indices, subscripts are flavour indices. \label{fig:4} }
\end{figure}

When the external fermions are SU(2) doublets,  for
instance   the first generation quark doublet  $q_1$, additional diagrams arise.
Firstly, there will be wavefunction corrections on  the
external doublet lines, and  all but the third  vertex diagram of figure
\ref{fig:3} will occur, but with the $W$ attached to the external
doublet line --- these diagrams all vanish.  In addition, there
will be diagrams, illustrated in figure \ref{fig:4}, where the
$W$ is exchanged between the external  fermion lines, and the
flavour-changing lepton lines.  These do not vanish, and correspond
to the one-loop diagrams  that renormalise and mix vector
four-fermion operators. 

The spinor contractions and momentum integral for the first two diagrams,
at zero external momentum, give a divergence 
\bea
-\frac{g^2}{4}   \frac{C}{\Lambda_{NP}^4}
 \frac{i }{16\pi^2 \epsilon}
\times 
(3+\xi)(\overline{u}_l \g^\a P_L u_k)
(\overline{u}_j  \g_\a P_L u_i) 
 \label{d4kW4f1}
\eea
whereas the last two diagrams give the cancelling term $\propto \xi$. 
It remains to perform the SU(2) contractions, that
define which operator mixes to which; these
can be read off the anomalous dimension matrices given in section
\ref{ssec:adm}.

 For the case where there
are identical fermions ($\ell_e$ as external fermions),
the operator basis is smaller (see eqn \ref{ellbasis}),
so the divergences due to $W$ exchange  among  fermions
look different. It is straightforward to check that
the same  divergences are generated
by operators that become identical in  the presence of
identical fermions.

\begin{figure}[htb]
\begin{center}
\epsfig{file=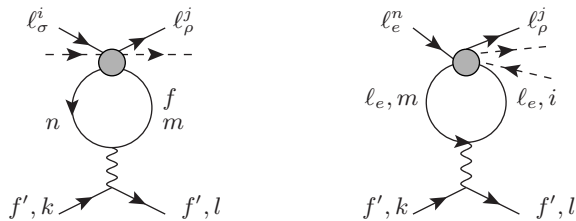,width=0.42\textwidth}
\end{center}
\vspace{-.5cm} 
\caption{$W$ penguin diagrams that occur when the external fermion
  is a doublet. The right penguin only occurs if
  the operator involves identical fermions, such
  as  two $\ell_e$ fields. \label{fig:penguins}}
\end{figure}

Finally, the $W$ bosons can mediate  penguin diagrams,
as illustrated in figure \ref{fig:penguins}.
For operators without identical fermions, only
the left penguin can occur, and  vanishes for 
${\cal O}_{NSI}$, ${\cal O}_{H2}$
and  ${\cal O}_{M2}$, due to a
trace over the SU(2) generator. 
 For $W$ penguins, there is
only a sum over the colour of quarks in the loop,
never a 2 for  tracing over SU(2) doublets, because the loop vanishes
as the trace of a generator in this case.
These diagrams  can change the  external fermion, {\it eg} $\ell_e \leftrightarrow q_1$, thereby mixing operators with  different
external fermions; for simplicity,  this mixing is  neglected
in the  RGEs of section \ref{ssec:adm}.
(It does not give additional constraints
when the external fermion is a quark doublet; 
it is interesting for external
lepton doublets and is briefly rediscussed
in section \ref{ssec:4.1}.)

 In  the  case of identical fermions
 (the external fermions are $\ell_e$, and  $\r$ or $\s$ is $e$),
 there could be two penguin  diagrams, due to the  identical fermions.  
However, since we consider vector operators,
which can be rearranged according to Fierz,
 the spinor  contractions  and momentum integrals
 for the two possible  diagrams  are 
 the same; only the SU(2) contractions can differ.
In particular, the
relative sign between the amplitudes is +, because the two
diagrams are  Fierz transformations of each other.

The  different SU(2) contractions for
 the two penguin diagrams should correspond
to the penguin contributions of two operators
which become identical when there are identical fermions.
For instance,  for external $q$, ${\cal O}_{M2}$ has no penguin
diagram, but ${\cal O}_{LQM2}$ generates
divergences $\propto 2{\cal O}_{LQM2}
-{\cal O}_{M2}$ via the penguin.  For the operators
with external $\ell_e$ and identical fermions, ${\cal O}_{M2}$
and ${\cal O}_{LQM2}$ are identical,  so the ``different''
SU(2) contraction that allows ${\cal O}_{M2}$ to have
a penguin diagram is just the SU(2) contraction that
allowed a penguin to ${\cal O}_{LQM2}$.
We conclude
that in the reduced basis   of operators with identical  leptons,
one must sum the penguin divergences  of the
different operators that become identical.

\subsection{The Higgs loops}

\begin{figure}[ht]
\begin{center}
\epsfig{file=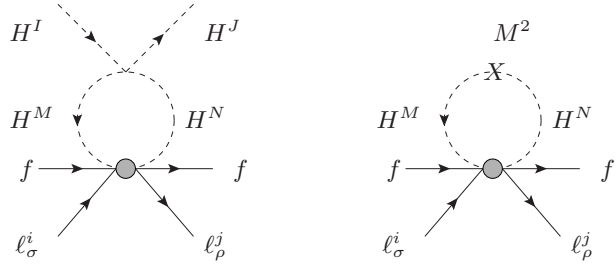,width=0.45\textwidth}
\end{center}
\vspace{-.3cm} 
\caption{ $H$ loops that mix and renormalise  dimension eight
  operators, and  mix them to dimension six via
  the  Higgs $M^2$ insertion. The external fermion is $f$\label{fig:5}.}
\end{figure}

Closing the Higgs legs and  inserting
$\lambda H^4$ can renormalise and
mix  the dimension eight operators.
Inserting instead $M^2$ on the scalar
line, as in the  right diagram of figure \ref{fig:5},
mixes the dimension eight operators
into    ${\cal O}_{M2}$ and   ${\cal O}_{LQM2}$.
These loops are straightforward to calculate, have
no subtleties in the presence of identical fermions, and
give rise to the anomalous dimensions given in
the following sections.

\subsection{Deriving RGEs }
\label{ssec:rge1loop}

We wish to obtain the  one-loop RGEs 
 for our operator coefficients, which, for a choice
 of  lepton flavour indices $\r,\s$, and external
 fermion $f$ are assembled 
 in a row vector
\beq
\vec{C} =  (C^{\r,\s}_{NSI,f}, C^{\r,\s}_{H2,f},..., C^{\r,\s}_{M2,f}) ~~~,
\label{vec}
\eeq
where ...  is the additional coefficients that could
arise if $f$ is an SU(2) doublet.  It is convenient,
  during  this derivation, 
  to multiply ${\cal O}_{M2}$ and ${\cal O}_{LQM2}$ by $M^2$,
  so that all the operators are of dimension 8.
  With this modification,
the  Lagrangian in $4-2\epsilon$ dimensions
can be expressed in terms of running fields and
parameters as
\bea
{\cal L} &=& ... + \frac{1}{\Lambda^4} \sum_f  \left\{  \vec{C}_A [Z]_{AB} \cdot
(Z_H^{n/2} Z_\ell  \mu^{(2+n)\epsilon}  \vec{O}_B) \right\}
\eea
where $n \in \{0,2\}$ is the number of Higgs legs
of the operator $O_B$. The bare coefficients  $\vec{C}_{bare} =  \vec{C} [Z] \mu^{(2+n)\epsilon} $ should
satisfy  $\frac{d}{d \mu} \vec{C}_{bare} = 0$,
which gives Renormalisation Group Equations
for the $C_A$s:
\bea
\mu \frac{\partial}{\partial \mu} {C}_A &=& -4\epsilon {C}_A
+ 2\epsilon (\vec{C}\cdot [Z])_{M2} \delta_{A,M2} -
\left( \vec{C}\cdot  \mu \frac{\partial g_i}{\partial \mu}  \frac{\partial [Z] }{\partial g_i} [Z]^{-1} \right)_A 
\label{rge1} \\
&=& \vec{C} \cdot [\Gamma] 
\eea
The operator $O_{M2}$ has dimension $8 - 4\epsilon$, whereas
 $O_{H2}$ and  $O_{NSI}$  are   $8 - 6\epsilon$-dimensional,
 which gives different ${\cal O}(\epsilon)$  terms in the
 RGEs. These terms give the anomalous dimensions mixing
  $O_{H2}$ and  $O_{NSI}$  to  $O_{M2}$, because the counterterms
in the M2 column of $[Z]$  are independent of $\lambda $ and $g_2$, so the last
  term vanishes. As a result, the  off-diagonal 
  anomalous dimensions,  as usual at one loop, are  twice
  the coefficient of $1/\epsilon$ in the counterterms.
For the diagonal anomalous dimensions, 
wavefunction contributions 
should be subtracted in the usual way
(because  the counterterms  for an amputated
operator are   represented by  $\vec{C}\cdot\ [C] =\vec{C}\cdot\ [Z] 
Z_H^{n/2} Z_\ell$, but we only want $[Z]$):
\bea
[\Gamma]_{AA} & =& 2 [C^{(1)}]_{AA} - 2Z^{(1)}_\ell - 2Z^{(1)}_H \delta_{1,n/2}  
\label{rges2} \\
{[\Gamma]}_{AB} & =& 2 [C^{(1)}]_{AB} ~~ ~,~ ~~~ A \neq B  \nonumber
\eea
where $Z^{(1)}$ is the coefficient of $1/\epsilon$ in $Z$.

Neglecting the running of the couplings ($g_2$,$y_t$, $\lambda$),
the solution is 
\bea
 \vec{C}(\mu_f)&=& \vec{C}(\mu_i) \cdot \left( [I]  + [\Gamma]
 \log \frac{\mu_f}{\mu_i}   + \frac{1}{2}[\Gamma \Gamma]
 \log^2 \frac{\mu_f}{\mu_i} + ...  \right)
\label{solnRGEs1l}
\eea
where, by analogy with running masses, the couplings in $[\gamma]$
are to be evaluated at $\mu_f$.

\subsection{The anomalous dimension matrix}
\label{ssec:adm}

For singlet external fermions,
in the basis $( C_{NSI}, C_{H2}, C_{M2} ) $,
 the anomalous dimension matrix is
\bea
[\Gamma] &=& \frac{g^2}{4\kappa}
\left[
\begin{array}{ccc}
-18  & 0 & 0\\
0 &-18   &  0\\
0&0&0\\
\end{array}\right]
+ 
\frac{1}{\kappa}
\left[
\begin{array}{ccc}
-4\lambda & 2\lambda &-2\eta\\
2\lambda & -4\lambda &  2\eta\\
0&0&0\\
\end{array}\right]
\nonumber\\
&&+  \frac{g^{'2}}{4\kappa}
\left[
\begin{array}{ccc}
-6 +24Y_f+ 16N_cY_f^2/3  & 0 & 0\\
0 &-6 +24Y_f+ 16N_cY_f^2/3  &  0\\
0&0&24Y_f+ 16N_cY_f^2/3\\
\end{array}\right]
~~ \label{ADfin}\\
{[\Gamma \Gamma]} &= &\frac{1}{\kappa^2}
\left[
\begin{array}{ccc}
d^2 + 4\lambda^2  & 4\lambda d
&   4\lambda\eta -2\eta (d  +d')\\
4\lambda d  & d^2+ 4\lambda^2  &
   -4\lambda\eta +2\eta (d +d') \\
0&0&d^{'2}\\
\end{array}\right] \nonumber
\eea
where $\kappa = 16\pi^2$, $\eta = M^2/\Lambda^2$, 
 and $d = -(9g^2/2 + 4\lambda +g^{'2}[1.5 -6Y_f-4N_{c,f} Y_f^2/3]) \sim -4$
is the diagonal anomalous dimension
of ${\cal O}_{NSI}$ and  ${\cal O}_{H2}$,
and $d'$  that of   ${\cal O}_{M2}$.

For doublet external fermions, in the basis
$( C_{NSI}, C_{H2}, (C_{CCLFV}+C_{CCLFV}^\dagger)/2, (C_{CCNSI}+C_{CCNSI}^\dagger)/2$,
$(C_{CCLFV}-C_{CCLFV}^\dagger)/2$, $ C_{LQM2},C_{M2})$, the anomalous dimension matrix 
is
\bea
[\Gamma] &=& 
-\frac{3 g^2}{\kappa} \left[
\begin{array}{ccccccccc}
\frac{5}{2} & 0 & 0 &-1 & 0 & 0 & 0 \\
 0 &\frac{5}{2} &-1 & 0 & 0 & 0 & 0 \\
 2 &-2 &\frac{3}{2} &-1 & 0 & 0 & 0 \\
-2 & 2 &-1 &\frac{3}{2} & 0 & 0 & 0 \\
 0 & 0 & 0 & 0 & \frac{5}{2} & 0 & 0 \\
0 & 0 & 0& 0 & 0 & 1  &-2  \\
 0 & 0  & 0& 0  & 0 &-2  & 1  \\
\end{array}\right]
\nonumber
\\
 &&+\frac{ g^2N_c}{3\kappa} \left[
\begin{array}{ccccccccc}
 0 & 0 & 0 & 0 & 0 & 0 & 0 \\
 0 & 0 & 0 & 0 & 0 & 0 & 0 \\
 0 &-2 & 2 & 0 & 0 & 0 & 0 \\
-2 & 0 & 0 & 2 & 0 & 0 & 0 \\
 0 & 0 & 0 & 0  & 2& 0 & 0 \\
0 & 0  & 0 & 0 & 0 & 2 &-1 \\
 0 & 0  & 0 & 0 & 0 & 0 & 0  \\
\end{array}\right]
+  \frac{1}{\kappa}
\left[
\begin{array}{ccccccccc}
-4\lambda & 2\lambda & 0 & 0 & 0 & 0 &-2\eta \\
 2\lambda &-4\lambda & 0 & 0 & 0 & 0 & 2\eta  \\
 0 & 0 &-4\lambda & 2\lambda & 0 & 4\eta & 0  \\
 0 & 0 & 2\lambda &-4\lambda& 0  &-4\eta & 0  \\
 0 & 0 & 0 & 0  &-2\lambda & 0 & 0 \\
 0 & 0 & 0 & 0 & 0 & 0 & 0 \\
 0 & 0 & 0 & 0 & 0 & 0 & 0 \\
\end{array}\right]
\label{AD2fin}
\eea
where  $\kappa = 16\pi^2$,   $\eta = M^2/\Lambda^2$,
and the first  matrix is from $W$ exchange,
the second is  the $W$ penguins and the last  is the Higgs.

In the case
with  external lepton doublets and identical fermions,
several operators are identical (see eqn \ref{redundancies}),
so  the anomalous dimension mixing operator A into  operator B is
the $\sum_{B'} \Gamma_{AB'}$ over  all the operators $\{B'\}$
who are identical to $B$. This rule applies to the second matrix of eqn(\ref{AD2fin}). Then for the penguins,  the rule is  to sum also  over 
the  identical operators in the column:
$ \Gamma_{AB} = \sum_{A',B'} \Gamma_{A'B'}$.
Then the anomalous dimension matrix, in the basis
 $( C_{NSI}, C_{H2},  C_{CCNSI+},C_{M2} ) $,
is
\bea
 [\Gamma] &=& 
-\frac{3g^2}{\kappa}  
\left[
\begin{array}{cccc}
\frac{5}{2}&0 &  -1 & 0\\
2 & \frac{1}{2} & -1 &  0\\
0 & 0 & \frac{1}{2}& 0\\
0&0&0&-1 \\
\end{array}\right]
+\frac{g^2N_c}{3\kappa}
\left[
\begin{array}{cccc}
1&0&  0 & 0\\
0 & 1 &  0 &  0\\
-5&1&4 &0 \\
0&0&0& 1 \\
\end{array}\right]
+\frac{1}{\kappa}
\left[
\begin{array}{cccc}
-4\lambda & 2\lambda&0 &-2\eta\\
2\lambda & -4\lambda &0&  2\eta\\
 -4\lambda& +4\lambda&- 2\lambda&-4\eta\\
0&0&0&0\\
\end{array}\right]
~~ \label{AD2finid}
\eea


\section{ Results}
\label{sec:results}

This section presents the LFV that is induced by  electroweak loop
corrections to NSI operators.
Section
\ref{ssec:LFV} summarises relevant experimental constraints on LFV,
then section  \ref{ssec:4.1} applies these constraints
to the  LFV coefficients induced by  loop corrections to NSI.  
Possible cancellations allowing to avoid these constraints
are discussed in section \ref{ssec:cancellations}.

\subsection{Experimental sensitivity  to  LFV operators}
\label{ssec:LFV}

Loop corrections to  NSI can induce
vector four-fermion operators (as given in eqn  \ref{belowmW}), that
  involve two charged leptons of different flavour, and
  two first generation fermions $e,u,$ or $d$. This
  section lists the experimental sensitivity to such
  coefficients. Since all the operators  considered here
  are hermitian (on doublet lepton flavour indices $\r\s$),
  we do not distinguish between bounds on $C^{\r\s ff}$ vs
   $C^{\s\r ff}$, and quote  bounds on only one.
  
If the lepton  flavours $\r,\s$ are $\mu$ and $e$, then  $\meee$
and $\mec$ are  sensitive to the LFV induced by loop corrections
to NSI operators.   Current bounds from
SINDRUM \cite{Bellgardt:1987du,Bertl:2006up}  at
  90$\%$ C.L. are
 $BR(\mu Au \to eAu) \leq 7.0 \times 10^{-13}$,
and  $BR(\meee)\leq 10^{-12}$, and give  sensitivities
(to the operator coefficients at $m_W$)
 \bea
 C^{\mu eee}_{V,LL}& \leq &7.8 \times 10^{-7} \\
 C^{\mu eee}_{V,LR}& \leq &9.3 \times 10^{-7} \\
 C^{\mu edd}_{V,LL}&\leq& 5.3 \times 10^{-8}\\
 C^{\mu edd}_{V,LR}&\leq& 5.4 \times 10^{-8}\\
C^{\mu euu}_{V,LL} &\leq& 6.0 \times 10^{-8}\\
C^{\mu euu}_{V,LR} &\leq& 6.3 \times 10^{-8}\label{muenumbers}
\eea
Experiments under construction
(COMET \cite{COMET},Mu2e \cite{Mu2e},Mu3e \cite{Mu3e}) will improve these
sensitivities by two orders of magnitude in a few years.

For one of  $\r,\s$ a $\tau$, and the other $\mu$  or $e$,
current bounds on $\tau \to \ell e^+e^-$  at
  90$\%$ C.L. give \cite{dim52}
 \bea
 C^{\tau eee}_{V,LL}& \leq &2.8 \times 10^{-4} \\
 C^{\tau eee}_{V,LR}& \leq &4.0 \times 10^{-4} \\
 C^{\tau \mu ee}_{V,LL}&\leq& 3.2 \times 10^{-4}\\
 C^{\tau \mu ee}_{V,LR}&\leq& 3.2 \times 10^{-4} ~~.
  \label{tauellnumbers}
\eea
These sensitivities again apply to the operator
coefficients at $m_W$.

The operators  with $u$  or $d$  quarks as  external fermions
can be probed by the LFV $\tau$ decays  $BR(\tau \to \{\mu,e\} \pi^0)
\leq \{1.1\times 10^{-7}, 8\times 10^{-8}\} $ \cite{tmp,tep},
$BR(\tau \to \{\mu,e\} \rho)\leq \{1.2\times10^{-8},1.8\times10^{-8}\}$
\cite{tlr} and
$BR(\tau \to \{\mu,e\} \eta)\leq \{6.5\times10^{-8},9.2\times10^{-8}\}$
\cite{tep} (all limits  at 90$\%$ C.L.). As noted in \cite{BHHS},
these three decays given complementary  constraints, because
the $\eta$ is an isospin singlet  ($\propto \bar{u} \Gamma u + \bar{d}
\Gamma d$) whereas the pion and $\rho$ are
isotriplets($\propto \bar{u} \Gamma u - \bar{d} \Gamma d$),
and  the decays to pions  or $\rho$s are respectively sensitive to
LFV operators involving the  axial or vector quark current.

It is convenient to normalise the pion decays to
the  SM  process $\tnp$ (with $BR(\tnp) = 0.108$\cite{PDB}), in order
to cancel the hadronic and phase space factors:
\bea
\frac{BR(\tlp)}{BR(\tnp)} = \frac{ |C^{\tau \ell uu}_{V,LR} -C^{\tau \ell uu}_{V,LL} -
  C^{\tau \ell dd}_{V,LR} +C^{\tau \ell dd}_{V,LL}|^2}{2|V_{ud}|^2}
\eea
where the $2$ is because
$\sqrt{2} \langle 0 |\overline{u}\g^\a \g_5u| \pi_0\rangle = 
\langle 0 |\overline{u}\g^\a \g_5d| \pi^-\rangle$. 
This gives
 \bea
 |C^{\tau euu}_{V,LR} -C^{\tau euu}_{V,LL}-C^{\tau edd}_{V,LR} +C^{\tau edd}_{V,LL}|& \leq &1.2 \times 10^{-3} \nonumber \\
 |C^{\tau \mu uu}_{V,LR} -C^{\tau \mu uu}_{V,LL}-C^{\tau \mu dd}_{V,LR} +C^{\tau \mu dd}_{V,LL}|& \leq &1.4 \times 10^{-3}
 ~~.
  \label{taupinumbers}
\eea
These  sensitivities apply  to the coefficients at the experimental scale
(not the weak scale as for eqns \ref{tauellnumbers} and \ref{muenumbers}).

The trick of normalising by an SM  decay is more subtle in the case
of $\tlr$, because
the $\rho$ decays to two pions, so the
$\tlr$  bounds are obtained by selecting a range of  $\pi^+\pi^-$
invariant-mass-squared
appropriate for  the $\rho(770)$. The corresponding SM decay is
$BR(\tnpp) =.255$, studied by Belle \cite{Belletnpp} over
a wide invariant-mass-squared. The fit to the  spectrum performed
by Belle suggests that $\sim 80\%$ of the events are due to
the $\rho(770)$, so for simplicity\footnote{A detailed fit and discussion of the form factors
  for $\tlpp$ is given in \cite{CPX}.} we suppose:
\bea
\frac{BR(\tlr)}{BR(\tnpp)} = \frac{  |C^{\tau \ell uu}_{V,LR} +C^{\tau \ell uu }_{V,LL}-
  C^{\tau \ell dd}_{V,LR} -C^{\tau \ell dd }_{V,LL}|^2}{2}
\eea
which
 gives
 \bea
 |C^{\tau euu}_{V,LR} +C^{\tau euu}_{V,LL}
 -C^{\tau edd}_{V,LR} -C^{\tau edd}_{V,LL}|& \leq &3.8 \times 10^{-4} \nonumber \\
 |C^{\tau \mu uu}_{V,LR} +C^{\tau \mu uu}_{V,LL}
 -C^{\tau \mu dd}_{V,LR} -C^{\tau \mu dd}_{V,LL}|& \leq &3.1\times 10^{-4}
 ~~.
  \label{taurhonumbers}
  \eea

  For the $\eta$,  we approximate   $f_\eta \simeq F_\pi \simeq 92$ MeV
  (see  \cite{feta} for a detailed discussion), so that
  \beq
  \frac{\Gamma(\tau \to \ell \eta)}{\Gamma(\tau \to  \nu \pi^-)} =
  \frac{ |C^{\tau \ell uu}_{V,XR} -C^{\tau \ell uu }_{V,XL}+
  C^{\tau \ell dd}_{V,XR} -C^{\tau \ell dd }_{V,XL}|^2}{2} ~~~,
  \eeq
and the current bounds on $\Gamma(\tau \to \ell \eta)$  imply
 \bea
 |C^{\tau euu}_{V,XR} -C^{\tau euu}_{V,XL}
 +C^{\tau edd}_{V,XR} -C^{\tau edd}_{V,XL}|& \leq &6.5 \times 10^{-4} \nonumber \\
 |C^{\tau \mu uu}_{V,XR} -C^{\tau \mu uu}_{V,XL}
 +C^{\tau \mu dd}_{V,XR} -C^{\tau \mu dd}_{V,XL}|& \leq &5.4\times 10^{-4}
 ~~.
  \label{tauetanumbers}
  \eea
  In coming years, Belle II could improve the sensitivity
to LFV $\tau$ decays by  one or two orders of magnitude \cite{Kou:2018nap}.
  
  For models that induce LFV on left-handed, or right-handed quarks, but not both,
  the bounds of eqns  (\ref{taurhonumbers}) and (\ref{tauetanumbers}) can be
  combined in a covariance matrix to obtain
 \bea
 |C^{\tau eqq}_{V,LX}|& \leq &7.1 \times 10^{-4} \nonumber \\
 |C^{\tau \mu qq}_{V,LX}|& \leq &5.9\times 10^{-4}
  \label{covarnumbers}
  \eea
where $q\in \{u,d\}$ and $X = L$ or $R$.

\subsection{LFV due to NSI}
\label{ssec:4.1}

We consider 
 combinations of operator
 coefficients which,  at tree level, 
 induce NSI but not LFV  (these were given
section \ref{sec:notn}), and     use
the RGEs obtained in  section
\ref{sec:loops} to estimate the effect of loops.
For example,  the one-loop [or two-loop]
mixing     of a given combination of  tree-level  coefficients,
can be obtained from the second [or third] term of
eqn (\ref{solnRGEs1l}),  with $\vec{C(\mu_i)}$ the  input (tree) coefficients
at the New Physics scale $\mu_i = \Lambda_{NP}$,
and $\vec{C(\mu_f)}$ the loop-induced combination
at the weak scale $m_W$.
 By matching  $\vec{C(\mu_f)}$ onto the low-energy theory, one
 obtains the LFV induced by the one-loop RGEs.

 The case of singlet external fermions is simple
 to discuss as an explicit example. Eqn (\ref{match})
 implies that NSI can arise at tree-level from 
$ C_{NSI}$  and/or $ C_{M2}$ (subdominant loop contributions
 to coefficients induced
 at tree level are neglected in the following.)
 For   only $ C_{NSI} (\Lambda_{NP}) \neq 0$,  
 eqn (\ref{ADfin}) gives
 \bea
 \Delta  C^{\r\s }_{H2,f}(m_W) &=& C^{\r\s }_{NSI,f} (\Lambda_{NP}) \times\left( \frac{2 \lambda}{(16\pi^2)}
 \log \frac{\Lambda_{NP}}{m_W} + \frac{4 \lambda d}{2(16\pi^2)^2}
 \log^2 \frac{\Lambda_{NP}}{m_W} +...\right)\nonumber\\
 \Delta  C^{\r\s }_{M2,f}(m_W) &=&  C_{NSI,f}^{\r\s } (\Lambda_{NP}) \times \left(- \frac{2 \eta}{(16\pi^2)}
 \log \frac{\Lambda_{NP}}{m_W} +\frac{4 \lambda \eta -2\eta(d+d')}{2(16\pi^2)^2}
 \log^2 \frac{\Lambda_{NP}}{m_W}  +...\right) \nonumber
 \eea
 where $d$ and $d'$ are defined after eqn (\ref{ADfin}).
 Matching onto the low-energy operators according
 to eqn (\ref{match}) with table \ref{tab:ops},
 gives, at first order in $1/(16\pi^2)$, a vanishing LFV coefficient
 $C_{V,LR}^{\r\s ff} = 0$, due to potential
 minimisation conditions.
 However, at second order in the one-loop
 RGEs,   ${\cal O}_{NSI}$ 
 induces LFV  at low energy:
 \bea
\Delta C_{V,LR}^{\r\s ff} =   \frac{C^{\r\s }_{NSI,f} (\Lambda_{NP})v^4}{\Lambda_{NP}^4}
 \frac{ 2\lambda(d-d') + 4\lambda^2 }{2(16\pi^2)^2}
   \log^2 \frac{\Lambda_{NP}}{m_W} \sim
10^{-4} \varepsilon_f 
 ~~~,\label{DC2l}
 \eea
 where $d-d'=-(9g^2/2 + 4 \lambda)$ if hypercharge is neglected,
  and for the  numerical
   estimates in this section, we conservatively take
   $\Lambda_{NP} \sim 250-300$ GeV in the logarithm.
 
 For $C_{M2}(\Lambda_{NP}) \neq 0$, the tree contribution to LFV must
 be cancelled by  $C_{H2}(\Lambda_{NP}) = - (\lambda/\eta) C_{M2}(\Lambda_{NP})$ 
 as given in eqn (\ref{vanishf}).
 Then the RGEs generate corrections to $C_{H2}$ and $C_{M2}$:
  \bea
 \Delta  C^{\r\s }_{H2,f}(m_W) &=& C^{\r\s }_{H2,f} (\Lambda_{NP}) \times \frac{ d}{(16\pi^2)}
 \log \frac{\Lambda_{NP}}{m_W} + ...\nonumber\\
 \Delta  C^{\r\s }_{M2,f}(m_W) &=&  C_{M2} (\Lambda_{NP}) \times  \frac{d'-2\lambda}{(16\pi^2)}
 \log \frac{\Lambda_{NP}}{m_W} 
 +... \nonumber
 \eea
 which match onto low-energy LFV  at one loop:
 \bea
\Delta C_{V,LR}^{\r\s ff} =   \frac{C_{M2} (\Lambda_{NP})v^2}{(16\pi^2) \Lambda_{NP}^2}
  \left[ -(d-d') -2\lambda   \right]
  \log \frac{\Lambda_{NP}}{m_W} \sim -2 \times 10^{-2}  \varepsilon_{f}
  \label{DC1l} ~~~.
  \eea
So a heavy New Physics model  that gives  NSI on 
singlet fermions  will  induce LFV via loops,
 which  is  the sum of eqns (\ref{DC1l})
 and (\ref{DC2l}).

 \begin{figure}[ht]
\begin{center}
\epsfig{file=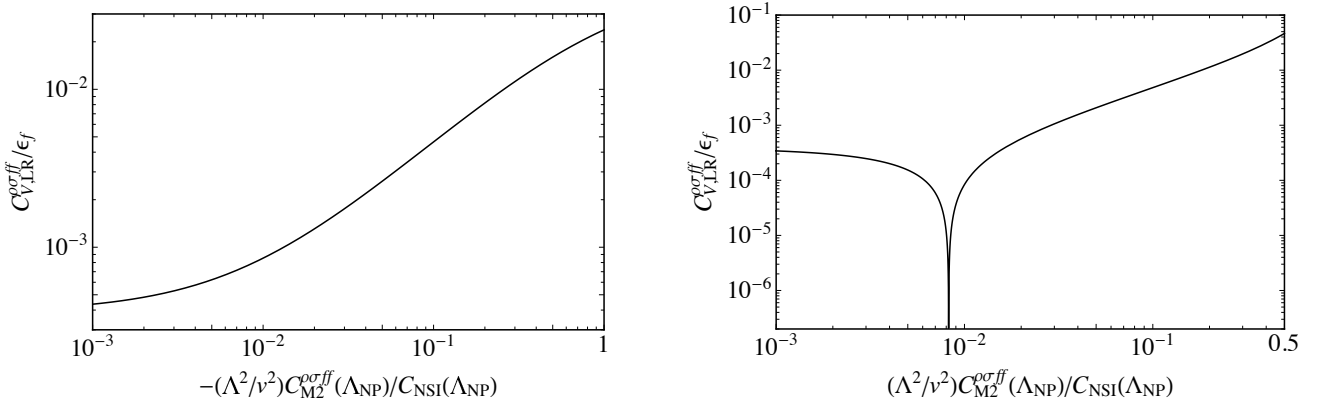,width=0.96 \textwidth}
\end{center}
\caption{
  The loop-induced LFV coefficient, normalised to the NSI coefficient
  $\varepsilon_f$, for SU(2) singlet external fermions $f$, as a function of the
  ratio of the two independent operator coefficients that can induce NSI: $C_{M2}(\Lambda_{NP})$ and $C_{NSI}(\Lambda_{NP})$.   $C_{H2}(\Lambda_{NP})$ is determined
  as a function of 
  $C_{M2}(\Lambda_{NP})$ by  the cancellation of  tree-LFV given in eqn~(\ref{vanishf}). The left plot is for negative   $C_{M2}/C_{NSI}$, and
 positive values   are in the plot to the right.
\label{fig:plot}}
\end{figure}

 In figures \ref{fig:plot}, the magnitude of
 the LFV coefficient is plotted against the ratio $C_{M2,f}\Lambda^2/C_{NSI,f} v^2$, for $\varepsilon_f = 1.0$ and assuming  tree-level LFV cancels
 according to eqn (\ref{vanishf}).
    For $|C_{M2,f}| > |C_{NSI,f} v^2/\Lambda^2|$,
   it is clear from eqns (\ref{match},\ref{DC1l},\ref{DC2l}) that
   $\varepsilon_f\simeq -C_{M2,f}v^2/\Lambda^2$, and 
   $ C_{V,LR} \simeq -2 \times 10^{-2}  \varepsilon_{f}$, so
   the plots illustrate the regions  $|C_{M2,f}| < |C_{NSI,f} v^2/\Lambda^2|$
   (For   $C_{M2,f} = C_{NSI,f} v^2/\Lambda^2$, $\varepsilon_f$
   vanishes  so $ C_{V,LR}/\varepsilon_f$ diverges.). 
   At  $C_{M2,f} = C_{NSI,f} v^2 \ln /(32 \pi^2 \Lambda^2)$,
   the figure shows 
   an ``accidental'' cancellation between the
    contributions
 to the LFV coefficient  from eqns (\ref{DC1l})
 and (\ref{DC2l}); we are reluctant to admit
 this loophole in the LFV constraints
 on NSI, because it is  difficult  to build models that  tune 
 Lagrangian parameters against logarithms of mass scales. 

 The  experimental  bounds on LFV from  section \ref{ssec:LFV} can now
 be applied to the loop-induced
 LFV coefficient,  obtained by  summing  eqns (\ref{DC1l})
 and (\ref{DC2l}).
 This gives  an upper bound on the NSI coefficient, that depends
 on the ratio $\Lambda^2 C_{M2}/(v^2C_{NSI})$ : $\varepsilon_f^{\r\s} \times$
 the value given in the plot must be smaller than the experimental
 constraint. For instance, for $C_{M2}(\Lambda) \lsim 10^{-2} C_{NSI}(\Lambda)$,
 $ \varepsilon_{f}^{\mu e}$ must be $< 10^{-3}\to 10^{-2}$ as given in
 the first column of table \ref{tab:bds2l},
 and  $\tau \leftrightarrow e,\mu$ NSI can be ${\cal O}(1)$.
 The $\tau$ decay bounds
are given in the second two columns of table  \ref{tab:bds2l}.
On the other hand, as soon as  
     $C_{M2} $ strays away from 0, the LFV
   bounds on NSI are more restrictive
(this is illustrated in figure \ref{fig:plot})
 --- 
     then  the LFV
 is ${\cal O} (10^{-2} \varepsilon_f)$, and the constraints on LFV
 are given in table \ref{tab:bds1l}. Notice however, that
all these estimates  are approximate because our EFT  calculation only
 allows to obtain the log$^n$-enhanced part of $n$-loop diagrams,
 and since the logarithm cannot be large, our results  should give
 the order of  magnitude, but not two significant figures.

  \renewcommand{\arraystretch}{1.5}
\begin{table}[h]
\centering
\begin{tabular}{|l|l|l|}
  \hline
$ \varepsilon_{e_R}^{\mu e} \lsim 9 \times 10^{-3}$&
 $\varepsilon_{e_R}^{\tau e} \lsim 4 $&
  $\varepsilon_{e_R}^{\tau \mu} \lsim  3$\\
\hline
  $ \varepsilon_{u_R}^{\mu e} \lsim 5 \times 10^{-4}$
&$\varepsilon_{u_R}^{\tau e} \lsim 7 $&
$  \varepsilon_{u_R}^{\tau \mu} \lsim  6$\\
\hline
$ \varepsilon_{d_R}^{\mu e} \lsim 6 \times 10^{-4}$&
$ \varepsilon_{d_R}^{\tau e} \lsim 7 $&
$  \varepsilon_{d_R}^{\tau \mu} \lsim  6 $\\  
\hline
\end{tabular}
\caption{ Bounds on flavour-changing NSI parameters from
  the non-observation of LFV processes among charged leptons,
  obtained from eqn (\ref{DC2l}) for
    SU(2) singlet external fermions. Comparable limits
    apply to  to the $\{\varepsilon_{fL}^{\rho\sigma}\}$ for doublets,
  as discussed after eqn (\ref{loopLFVq}). 
  These bounds, which are almost unavoidable,
  arise from   two-loop contributions (${\cal O}(\alpha^2 \log^2 $))
  of the NSI operators to
  LFV processes. 
\label{tab:bds2l}}
\end{table}
   \renewcommand{\arraystretch}{1}

\renewcommand{\arraystretch}{1.5}
  \begin{table}[h]
\centering
\begin{tabular}{|l|l|l|}
  \hline
  $ \varepsilon_{e_R}^{\mu e} \lsim 5 \times 10^{-5}$&
  $ \varepsilon_{e_R}^{\tau e} \lsim 2 \times 10^{-2}$&
  $  \varepsilon_{e_R}^{\tau \mu} \lsim  2\times 10^{-2}$\\
  \hline
  $ \varepsilon_{u_R}^{\mu e} \lsim 3 \times 10^{-6}$&
  $\varepsilon_{u_R}^{\tau e} \lsim 4\times 10^{-2} $&
$  \varepsilon_{u_R}^{\tau \mu} \lsim  3\times 10^{-2}$\\
  $ \varepsilon_{d_R}^{\mu e} \lsim 3 \times 10^{-6} $&
  $ \varepsilon_{d_R}^{\tau e} \lsim 4\times 10^{-2} $&
$ \varepsilon_{d_R}^{\tau \mu} \lsim  3 \times 10^{-2}$\\
 \hline
\end{tabular}
\caption{ Bounds on flavour-changing NSI parameters from
  the non-observation of LFV processes among charged leptons,
 obtained from eqn (\ref{DC1l}) for NSI on SU(2) singlet
external fermions.
  Comparable limits
    apply to  to the $\{\varepsilon_{fL}^{\rho\sigma}\}$ for doublets,
  as discussed after eqn (\ref{loopLFVq}).
  These bounds arise from   one-loop contributions
  (${\cal O}(\alpha \log $)) of the NSI operators to
  LFV processes, and 
  can be avoided  in models 
   that generate particular
  patterns of coefficients  as discussed in the text.
\label{tab:bds1l}}
\end{table}
 \renewcommand{\arraystretch}{1}

If the external fermion is an SU(2) doublet, the situation is more involved.  
 It is again the case
that $C_{NSI}$ first mixes into LFV 
at ${\cal O}(\alpha^2 \log^2 $), but for external doublet quarks,
the other five  coefficients all
induce LFV at ${\cal O}(\alpha \log $) .  In order to avoid
tree-level LFV, those  five coefficients must satisfy two constraints,
obtained by setting eqns (\ref{vanishq}) to zero. Then 
they will induce LFV as given by the RGEs of eqn
(\ref{AD2fin}):
\bea
\Delta C^{\r\s uu}_{V,LL} &=& \frac{v^2}{\Lambda^2}\frac{\log(\Lambda/m_W)}{16\pi^2}
\left([\frac{9}{2} g^2 +2\lambda] C^{\r\s}_{M2,q}
-6g^2\frac{v^2}{\Lambda^2} C^{\r\s}_{CCNSI+,q}
+g^2 C^{\r\s}_{LQM2,q}\right)
\nonumber\\
\Delta C^{\r\s dd}_{V,LL} &=& \frac{v^2}{\Lambda^2}\frac{\log(\Lambda/m_W)}{16\pi^2}
\left( C^{\r\s}_{M2,q}+ C^{\r\s}_{LQM2,q}\right) [\frac{9}{2} g^2 +2\lambda]  
\label{loopLFVq}
\eea
If  NSI are due to some  subset of 
$C_{CCNSI+,q}$, $C_{M2,q}$ and $C_{LQM2,q}$,
and the LFV coefficients of eqn (\ref{loopLFVq}) do not vanish,
then the bounds of table \ref{tab:bds1l}
would generically apply. (We
do not make plots in this case, because
there are four independent coefficients).

On the other hand, the above equations
contain three coefficients, so it is  possible
for the New Physics model to arrange them 
such that  the  ${\cal O}(\alpha \log) $  LFV on $u_L$ and $d_L$
currents vanishes:   the coefficients $C_{H2,q}$, $C_{CCLFV+,q}$,
$C_{CCNSI+,q}$, $C_{M2,q}$ and $C_{LQM2,q}$ must all be non-zero,
and satisfy the four relations obtained by setting 
eqns (\ref{loopLFVq}) and (\ref{vanishq}) to vanish.
If a model could be constructed to implement this
cancellation, it is possible that there would be
not-log-enhanced one-loop contributions to LFV operators;
however,
to verify that in EFT would require
 going beyond our leading-log analysis.
It is however sure, from our one-loop RGEs,  that
LFV will  be induced at  ${\cal O}(\alpha^2 \log^2) $,
so that constraints of order those in table  \ref{tab:bds2l} would apply.
As in the
case of external SU(2)-singlet fermions, these constraints
also apply if  the model matches only onto ${\cal O}_{NSI,q}$
at the scale $\Lambda$, with all the other coefficients
relatively suppressed by $\sim 10^{-2}$. The
exact formulae for these  ${\cal O}(\alpha^2 \log^2) $
contributions  are straightforward to obtain from the
third term in eqn (\ref{solnRGEs1l}); they are
not quoted here because they are lengthy.

It is interesting to resurrect the
``external-fermion-changing'' $W$-penguin diagrams of  figure
\ref{fig:penguins}, before giving results for the case
where the external  fermion is a lepton doublet.
These penguins can  change the external fermion
$\ell_e \leftrightarrow q$, so,  for instance,
an operator with external $\ell_e$
could generate
one-loop LFV on $u_L$ and $d_L$.
Requiring that the model choose its parameters
to cancel this LFV gives an additional constraint
on NSI for doublet leptons when
$\r\s \in \{\mu,\tau\}$
that is given in eqn (\ref{qpenguin}).

For external  $\ell_e$,  the NSI
and LFV are different if one of $\r,\s$ is
first generation.
When yes,   tree level  NSI and LFV are respectively
  generated  by the coefficient combinations given
in eqns~(\ref{NSIlid}) and (\ref{vanishf}).
For  $\r,\s \in \{\mu,\tau\}$,
the  combinations are given in eqns (\ref{vanishl})
and (\ref{nsile}). In the following,
we suppose that the tree-LFV combinations
of eqns (\ref{vanishf}) and (\ref{vanishl}) vanish.

The operator $ {\cal O}_{NSI,\ell}$, which contributes
to tree-level NSI,  first induces LFV 
at  ${\cal O}(\alpha^2 \log^2) $.
NSI can also arise due to $ C_{M2,\ell}$,
 in which case the  one-loop LFV is  
different depending if one of $\r,\s$ is
first generation. 
When yes, then
the one-loop LFV  on electrons is:
\bea
\Delta C^{\r\s ee}_{V,LL} &=&  \frac{v^2}{\Lambda^2}\frac{\log(\Lambda/m_W)}{16\pi^2}
\left([\frac{15}{2}g^2 +2\lambda]  C^{\r\s}_{M2,\ell} + \frac{g^2}{3} C^{\r\s}_{CCNSI+,\ell} \right) ~~~,
\label{loopLFVlid}
\eea
and the $W$-penguin-induced LFV on
quarks vanishes when  eqn (\ref{vanishf}) does.
So if NSI are induced by
$C_{M2,\ell}$, then the model can tune coefficients
to cancel tree and one-loop  LFV, by ensuring
that eqns (\ref{vanishf}) and (\ref{loopLFVlid}) vanish.

For $\rho$ and $\s \in \{\mu,\tau\}$,
the one-loop LFV is induced on $u_L$ and
$d_L$ by the $W$ penguins
\bea
\Delta C^{\r\s uu}_{V,LL} &=&
\frac{g^2}{3}\frac{v^2}{\Lambda^2}\frac{\log(\Lambda/m_W)}{16\pi^2}
\left(\frac{\eta}{\lambda}C^{\r\s}_{H2,\ell} + C^{\r\s}_{M2,\ell}\right)
\nonumber\\
\Delta C^{\r\s dd}_{V,LL} &=&
\frac{g^2}{3}\frac{v^2}{\Lambda^2}\frac{\log(\Lambda/m_W)}{16\pi^2}
\left(2\frac{\eta}{\lambda}C^{\r\s}_{CCLFV+,\ell} + C^{\r\s}_{LQM2,\ell}\right)
\label{qpenguin}
\eea
and on leptons:
\bea
\Delta C^{\r\s ee}_{V,LL} &=&  \frac{v^2}{\Lambda^2}\frac{\log(\Lambda/m_W)}{16\pi^2}
\left([\frac{9}{2}g^2 + 2\lambda]  (C^{\r\s}_{M2,\ell} + C^{\r\s}_{LQM2,\ell})
+\frac{g^2}{3}  C^{\r\s}_{LQM2,\ell}  +\frac{2}{3}g^2 \frac{v^2}{\Lambda^2} C^{\r\s}_{CCLFV+,\ell} \right)
\label{loopLFVl}
\eea
So  if   NSI arise due to an operator other than ${\cal O}_{NSI}$, then  
at least two  coefficients must be  cancel against each other to avoid
tree LFV(as shown in eqn \ref{vanishl}), and LFV will arise at one loop  unless
 the  model arranges eqns (\ref{loopLFVl},\ref{qpenguin}) to vanish. 

 In summary, for external lepton doublets, the LFV constraints
 are similar the case of an external quark doublet: generically,
 the bounds of table \ref{tab:bds1l} would apply; in the case where the
 model  matches only onto ${\cal O}_{NSI}$, or where
 it arranges its coefficients to cancel the LFV at ${\cal O}(\a\log)$,
 then the bounds of \ref{tab:bds2l} would apply.

\subsection{Cancellations}
\label{ssec:cancellations}

The results given in tables \ref{tab:bds1l}  and \ref{tab:bds2l}
are  not in reality  ``bounds'' on
NSI from LFV processes, but rather  ``sensitivities'':
NSI coefficients larger than  the given value 
{\it could} mediate  LFV  rates above the experimental limit,
but not necessarily,  in the case where 
their contribution to LFV is cancelled by  other coefficients.
This section lists some possible cancellations
that could allow NSI to evade the LFV constraints.

\ben
\item As already  discussed, for external fermions that are SU(2) doublets,
  there  are enough operators such  that, not only  the combination
  of coefficients  which contributes at  tree  level to LFV
  can be chosen to vanish, but  also the coefficient combination
  that contributes at $\alpha\log$.
  But the  two-loop ${\cal O}(\alpha^2 \log^2)$  bounds of  table \ref{tab:bds2l} would still apply.

\item  We neglected possible  cancellations  between
  flavours or chiralities of quarks\footnote{The experimental bounds
    on leptonic decays
    constrain individually the coefficients
    of different chirality.} in the
  experimental  sensitivities of section \ref{ssec:LFV}.

  In the case of NSI involving $\tau\leftrightarrow \ell$ flavour
  change, the $\tau$ decay bounds  quoted   do not
  constrain the  isosinglet vector combination
  $C_{V,XL}^{\tau \ell uu} + C_{V,XR}^{\tau \ell uu}
  +C_{V,XL}^{\tau \ell dd} + C_{V,XR}^{\tau \ell dd}$. 
  The authors are unaware of restrictive bounds
  on this combination; if indeed they are absent,
  then  tree   LFV bounds for $\tau \leftrightarrow \ell$
  NSI would not apply to
  an  NSI model
  where the low-energy LFV coefficients are equal
 for  external fermions $f = q_L,u_R,d_R$.
  This  equality could substitute for imposing the
  tree cancellations  of eqn (\ref{vanishq}).
  However, the coefficients  of operators with
  external fermions $q$,$u_R$ and $d_R$ all run
  differently  (the last two due to different hypercharge),
  so LFV would still arise at one loop, and the one-loop
  bounds would apply, unless further cancellations
  are arranged.
  
  In the case of  $\mu\leftrightarrow e$ NSI,
  the $\mu \to e$  conversion bounds 
apply to a weighted sum of the $u$ and $d$
vector  currents, where the weighting factor depends
on the target nucleus. It is not
possible to  avoid the  bound by
cancelling $u$ vs $d$ coefficients, because there
are restrictive bounds on $\mu \to e$
conversion on Gold (Z=79, used to
obtain  eqn \ref{muenumbers}) and Titanium
(Z=22, $BR( \mu Ti \to e Ti) \leq 4.2\times 10^{-12}$),
which have different $n/p$ ratios,
so together constrain  the $u-d$ combination
a factor of 2 less well than
 $u+d$. However,  the sensitivity of $\mu \to e$ conversion to
  the axial vector LFV operator $(\overline{e}\g^\a P_L \mu)
  (\overline{q} \g_\a \g_5 q)$, is $\sim$ three orders of magnitude
  weaker (below $m_W$,
  the axial vector  mixes via  the RGEs of QED to the vector operator).
  So if loop corrections to NSI
 generated LFV on the axial quark current, the LFV bound on NSI would be
 weakened by $10^3$.  
 
 This requires NSI on doublet and singlet quarks
 (involving operators other than ${\cal O}_{NSI}$),
 whose coefficients satisfy the zero-tree-LFV conditions,
 and where the external doublet  coefficients
 are of comparable magnitude  and  opposite sign 
 to the singlet coefficients.  Then U(1) and
 SU(2) penguin diagrams,
 that could mix these operators to  those with external
 electrons, vanish due to the zero-tree-LFV condition,
 and the bounds in the second and third row of
 the first column of table
 \ref{tab:bds2l} could be relaxed by three orders of magnitude.

\item  We neglected  the possibility that the  model induces ``other'' LFV
  not included  in our subset of operators
 (for instance, tensor or scalar  four-fermion operators),
  that could mix
  into it and  cause cancellations at low energy.
  
\item 
  We do  not allow cancellations  between
  Wilson coefficients at $\Lambda$
  (expressed in terms of parameters
  of the high-scale theory), against other
  Wilson coefficients multiplied by log$(v/\Lambda)$,
  because this would be  ``unnatural'' in EFT
  (In principle, the model predicts
  the couplings,  but   the observer  chooses the
  scale at which experiments are done, and therefore
  the  ratio in the log.).  
  However,  such ``accidental'' cancellations can occur and
  be numerically important; an example would be a model whose coefficients sit in the valley of figure
  \ref{fig:plot}.
  
  \een

\section{Discussion/Summary}
\label{sum}

 We  consider  New Physics models whose mass scale $\Lambda$ is  above
  $m_W$,  that induce  neutral current, lepton flavour-changing Non Standard neutrino Interactions (see eqn \ref{eqn1}), referred to as NSI.
  In Effective Field Theory (EFT), we study the Lepton Flavour Violating
  (LFV)  interactions that  such models can induce both at tree level, and   due to
  electroweak loop corrections.

  Section \ref{sec:notn} discusses the operator bases for
  the two  EFTs  used in this manuscript. Above 
  the weak scale  is the   $SU(3) \times SU(2)\times U(1)$-invariant
  SMEFT with dynamical Higgs and $W$-bosons, and below
  $m_W$ is a QED$\times$QCD-invariant theory where
  NSI cannot mix to LFV. 
  The dimension six and eight operators that we use above
  $m_W$ are given in eqns
  (\ref{opsd6us}) and (\ref{opsSU2}),   and their matching onto low-energy
  NSI, LFV and Charged Current operators is given in table \ref{tab:ops}.
  We refer to the not-$\nu$ fermions of the interaction as ``external''
  fermions;  if these are SU(2) singlets, the
  operator basis above $m_W$  contains only three operators. 
  The additional operators required for  external
    doublet quarks or leptons  are discussed in
  section \ref{ssec:2.1} and appendix \ref{app:basis}.

  We require that at tree level, the  model induces
  only NSI or Charged Current interactions, so the coefficients of LFV operators are required to vanish. The
  coefficients of low-energy LFV operators,
  induced at tree level by the operators from above $m_W$,
  are given in section \ref{ssec:<mW}, for
  the various possible external fermions. They  vanish
   if the model only matches onto the operators ${\cal O}_{NSI}$
  or  ${\cal O}_{CCNSI+}$ at $\Lambda$, or if there are
  cancellations among the coefficients of other operators,
  as given in  section \ref{ssec:<mW}. We allow
  arbitrary cancellations among  coefficients of  four-fermion operators
  of dimension six and eight, because such cancellations
  are natural in the Standard Model, where the potential
  minimisation condition  $-M  + \lambda v^2=0$ relates
  operators of different dimension and different number
  of Higgs legs.

  Section \ref{sec:loops} calculates  one-loop
  Renormalisation Group Equations (RGEs) for the operators above
  $m_W$.  These one-loop RGEs encode the $W$ and Higgs-induced mixing between
  NSI and LFV operators. The SU(2) gauge interactions ($\propto g_2 \sim 2/3$)
  and Higgs self-interactions ($\propto \lambda \sim 1/2$) are included;
  Yukawa couplings are neglected because they are  small for the
  external fermions which are first generation, and  hypercharge is neglected
  because it does not change the  SU(2) structure of the operators.

 The EFT  performed here is an expansion
in $\alpha^n \log^{n-m}$, where the one-loop RGEs
give the $m=0$ terms for all $n$, the two-loop
  RGEs would give  the $m=1$ terms for all $n$, and so on.
This differs from model calculations, which are
  usually expansions in  the number of loops or
in $\alpha^m$. The EFT expansion  gives a numerically
  reliable result 
when the logarithm is large, 
being the numerically
dominant term at each order in  $\alpha$.
 In the case of NSI models studied here, the log is not large, so may
  not be the only numerically relevant loop contribution to LFV
  in a particular model. (Appendix \ref{app:matching}
  discusses additional log-enhanced contributions to  the mixing of
  NSI to LFV that arise from using one-loop minimisation conditions
  for the Higgs potential.)

However,  in this study, we are
interested in  the $(\alpha \log)^n$  terms for
three reasons: firstly, they
are ``model-independent'', meaning we
can calculate them in EFT  and they
arise in all heavy New Physics models.
Second, they are independent of
the renormalisation scheme introduced for
the operators in the EFT. This is important,
because there are no operators in a renormalisable
high-scale model, so results that  depend on
the  operator renormalisation scheme 
can not be a prediction of the model.
Thirdly, the $\log \Lambda/m_W$ terms
are interesting because it is not obvious to cancel a log
against non-logarithmic contributions. So we anticipate
that the logs give a reliable model-independent
estimate of the size, or loop order, of
the LFV induced in models that give NSI. 

Section \ref{sec:loops} calculates the one-loop anomalous dimensions
for the three relevant  cases: external fermions which are
SU(2) singlets ($e_R,u_R$ and $d_R$), SU(2) doublets that
are not identical to   the lepton doublets
participating in the NSI (so doublet quarks $q$, and
$\ell_e$ when the NSI involve $\ell_\tau$ and $\ell_\mu$),
and finally external fermions which are lepton
doublets $\ell_e$ when the NSI current involves
$\ell_e$. The anomalous dimension matrices are
respectively given in eqns (\ref{ADfin}),(\ref{AD2fin})
and (\ref{AD2finid}).

An estimate for low-energy LFV
can  be obtained by matching the New Physics model
onto a vector  of operator coefficients at $\Lambda$,
which is input  as $\vec{C}(\mu_i)$ into the solution of the RGEs given
in eqn (\ref{solnRGEs1l}), with the appropriate
anomalous dimension matrices from section \ref{sec:loops}.
The output vector of this
equation, $\vec{C}(m_W)$, gives the coefficients
that can then be matching onto the LFV operators
below $m_W$ according to table \ref{tab:ops}.
This is performed in section \ref{ssec:4.1}.
The example of SU(2)-singlet external fermions
is discussed in some detail because this
case has the fewest free parameters; a
reader  with a different
selection of  operator coefficients can easily
calculate the  one-loop LFV  from the
results in section \ref{ssec:4.1},
and the two-loop LFV from eqn  (\ref{solnRGEs1l}).
The predicted LFV can then be compared
to current constraints on LFV that are
listed in section \ref{ssec:LFV}.
  
 In this manuscript, we allow arbitrary cancellations
among coefficients at each order in the $\ln/(16\pi^2)$
expansion, but neglect possible cancellations between orders.
This is discussed in section \ref{ssec:cancellations}.
So we require low-energy LFV to cancel at tree level,  then
enquire if it is induced at one or two loop, and examine
whether the coefficients can be  chosen to cancel
the loop-induced LFV. 
We find that almost all the operator  combinations
which at tree level  match  onto NSI without generating  LFV,
  will generate LFV at one loop, 
  suppressed with respect to NSI
  by a factor ${\cal O}( \log/(16\pi^2)) \sim 10^{-2}$.
  So generically,  NSI should satisfy the
  bounds given in table \ref{tab:bds1l}:
  $\varepsilon_f^{\mu e} \lsim 10^{-4}\to 10^{-5}$,
  $\varepsilon_f^{\tau \ell} \lsim 10^{-1}$.
  However, 
  there is one  dimension eight operator,  ${\cal O}_{NSI}$,
  for which  the log-enhanced one-loop LFV
  vanishes. Also, for
  external doublet fermions,  there are
  enough operators that it could be possible
  to arrange the coefficients  to cancel
  the log-enhanced part of the one-loop
  contribution to LFV. In  both these cases\footnote{ In the opinion of the authors of this manuscript,
  it could be interesting to build a model that
  induces only ${\cal O}_{NSI}$, or implements
  the appropriate cancellations  among
  operator coefficients. One could then check
  whether the complete one-loop contribution
  to LFV vanishes, or only the log-enhanced part.},
  LFV is generated at two-loop, so
  suppressed  by a factor ${\cal O}(\a^2 \log^2) \sim 10^{-4}$,
  and  NSI should satisfy the bounds of
  table \ref{tab:bds2l}:
  $\varepsilon_f^{\mu e} \lsim 10^{-2}$,
  $\varepsilon_f^{\tau \ell} \lsim$ few.
  Some other cancellations that could allow  NSI to be
  compatible with the LFV bounds are briefly
  discussed in section \ref{ssec:cancellations}.

\subsubsection*{Acknowledgements}

We thank Gino Isidori for motivation to begin this work.
MG is supported in part by the UK STFC under Consolidated Grant ST/L000431/1
and also acknowledges support from COST Action CA16201 PARTICLEFACE.

\appendix


\section{Identities and SM  Feynman rules}
\label{sec:ids}

\begin{figure}[ht]
\begin{center}
\epsfig{file=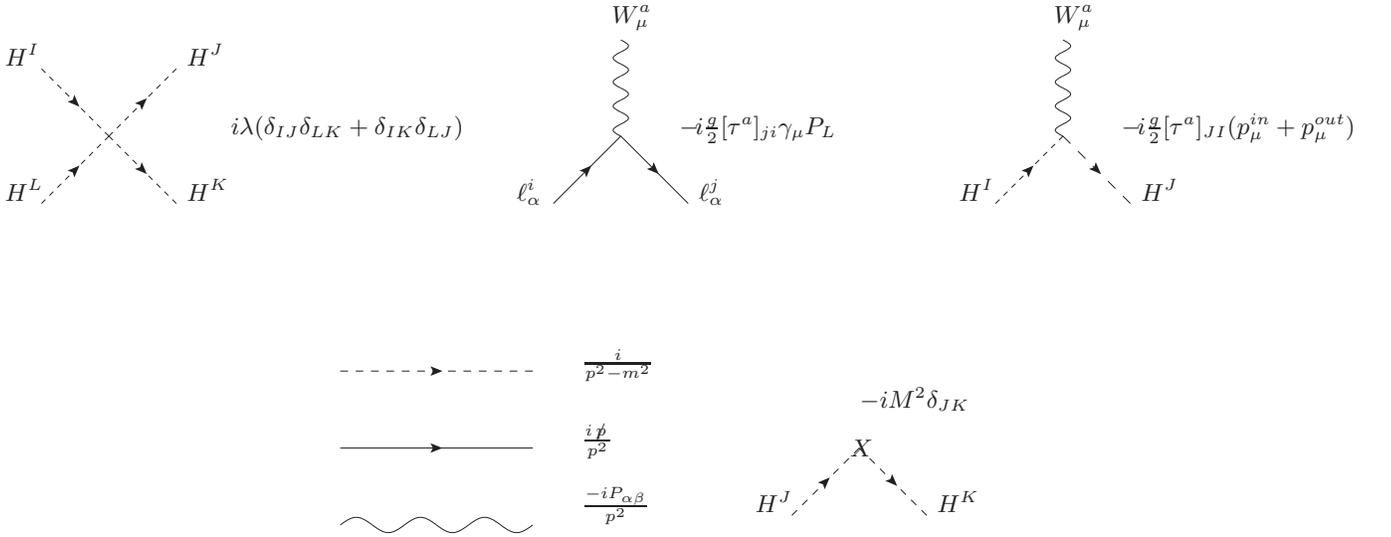,width=\textwidth}
\end{center}
\vspace{-.3cm} 
\caption{ Feynman rules for dimension-four interactions. \label{fig:1}
For the gauge boson propagator $P_{\a\b} =g_{\a\b} + (\xi -1)k_\a k_\b/k^2$. }
\end{figure}

The Pauli matrices and antisymmetric $\epsilon$ are
\bea \epsilon  =  \left[
\begin{array}{cc}
0&1\\
-1 &0
\end{array}
\right] ~~,~~\vec{\tau} = \left(\left[
\begin{array}{cc}
0&1\\
1 &0
\end{array}
\right],  \left[
\begin{array}{cc}
0&-i\\
i &0
\end{array}
\right] ,  \left[
\begin{array}{cc}
1&0\\
0 &-1
\end{array}
\right]
\right)~~~.
\label{eqnApp1}
\eea
The following identities  are useful:
\bea
2 \epsilon_{iI} \epsilon_{jJ} &= &\d_{ij} \d_{IJ} -\tau_{ij}^a \tau_{a,IJ} ~~{\rm Fierz}\label{Fierz}\\
\frac{1}{4}\tau_{ij}^a \tau_{a,kl} &=& \frac{1}{2}\d_{il} \d_{kj} - \frac{1}{4}\d_{ij} \d_{kl}  ~~~~{\rm SU(N)}
\nonumber
\\
\epsilon_{iJ}\epsilon_{kJ}&=&\delta_{ik}
\nonumber
\eea
where the first two   imply 
\bea
\epsilon_{ij}\epsilon_{kl} &=& \delta_{ik}\delta_{jl}-\delta_{il}\delta_{jk}.
\label{useful1}
\eea

\section{Dimension eight four-fermion operators}
\label{app:basis}

\subsection{constructing all possible SU(2) contractions}

The aim is to build all  possible SU(2) contractions for
an operator constructed from the fields:
\beq
 (\overline{\ell}^i_\r   \g_\a   \ell^j_\s)(\overline{q}^k  \g^\a q^l )  (H^{\dagger M}    H^N)
\label{fields}
\eeq
where $\{i,j,k,l,M,N\}$ are SU(2) indices.
For $R$  in the doublet representation of SU(2), invariants
can be constructed as follows:
$$
R^\dagger R ~~,~~
R \epsilon R  ~~,~~
R^* \epsilon R^* ~~,~~
R^\dagger \tau^a R R^\dagger \tau^a R ~~~,~~
\varepsilon_{abc} R^\dagger \tau^a R R^\dagger \tau^b R  R^\dagger \tau^c R~~.
$$

Consider first the $\tau\tau\tau$ contraction. Multiplying the product of
two Pauli matrices   by $\sum_{a,b} \tau^a \tau^b$ gives:
\bea
\sum_{a,b} \tau_{ij}^a \tau_{kl}^b {\Big (} \tau^a_{MR} \tau^b_{RN}{\Big ) } =
\sum_{a,b} \tau_{ij}^a \tau_{kl}^b {\Big (}
\delta^{ab}\delta_{MN} +  \sum_c i \varepsilon^{abc} \tau^c_{MN}{\Big ) } 
\eea
and using the identities of eqn (\ref{Fierz}),  allows
to write:
\bea
i \varepsilon^{abc} \tau_{ij}^a \tau_{kl}^b \tau^c_{mn} &=&
2 \d _{il}\d_{Mj} \d_{kN} -\d _{ij}\d_{Ml} \d_{kN} -\d _{iN} \d_{kl} \d_{Mj} 
-\d _{il} \d_{kj} \d_{MN} + \d _{ij} \d_{kl} \d_{MN}  
\eea
so this operator can be exchanged for $\d\d\d$ contractions. 
The   $\tau\tau$, and  $ \epsilon \epsilon$ contractions
can be rewritten  as $\d\d$s using the Fierz and SU(2)
identities of eqn (\ref{Fierz}),
so a complete set of operators is the
inequivalent $\d\d\d$ contractions.

There are  six possible  $\d\d\d$ contractions
(the permutations of three objects)
for the fields of eqn (\ref{fields}):
\bea
\d _{ij}\d_{kl} \d_{MN} &&~~~~~~~~~~~~~~~~~~~~~~~~~~~~~~\to  {\cal O}_S \nonumber\\
&=& \d_{kl}(- \epsilon_{iM}  \epsilon_{Nj} + \d_{kN} \d_{Ml}) \to    {\cal O}_{H2} -  {\cal O}_{NSI}   \nonumber\\
\d _{il}\d_{kj} \d_{MN} &=& \frac{1}{2}(\d _{ij}\d_{kl} + \tau^a _{ij} \tau^a _{kl} )
\d_{MN} \to  \frac{1}{2} {\cal O}_S  + \frac{1}{2}  {\cal O}_{TLQ}  \nonumber\\
&=&\frac{1}{2}{\Big \{} \d _{il} ( \d_{kN} \d_{Mj} -  \epsilon_{kM}  \epsilon_{Nj}  ) +  \d _{kj} (  \d_{iN} \d_{Ml} -  \epsilon_{iM}  \epsilon_{Nl} ) {\Big \}} \to \frac{1}{2}(  {\cal O}_{CCLFV} +{\cal O}_{CCLFV}^\dagger - {\cal O}_{CCNSI} -   {\cal O}_{CCNSI}^\dagger )\nonumber \\
\d _{iN}\d_{kl} \d_{Mj} &&~~~~~~~~~~~~~~~~~~~~~~~~~~~~~~  \to   {\cal O}_{H2}  \nonumber\\
&=& \frac{1}{2}(\d _{ij}\d_{MN} + \tau^a _{ij} \tau^a _{MN} )
\d_{kl} \to  \frac{1}{2} {\cal O}_S  + \frac{1}{2}  {\cal O}_{TLH}   \nonumber\\
\d _{ij}\d_{kN} \d_{Ml} &=& \frac{1}{2}(\d _{kl}\d_{MN} + \tau^a _{kl} \tau^a _{MN} )
\d_{ij} \to  \frac{1}{2} {\cal O}_S  + \frac{1}{2}  {\cal O}_{TQH}  \nonumber\\
\d _{il}\d_{kN} \d_{Mj}
&&~~~~~~~~~~~~~~~~~~~~~~~~~~~~~~  \to  {\cal O}_{CCLFV}
\nonumber\\
 \d _{iN}\d_{kj} \d_{Ml} &&~~~~~~~~~~~~~~~~~~~~~~~~~~~~~~  \to  {\cal O}_{CCLFV}^\dagger
 \label{dddcontractions}
 \eea
 where after the arrows, the contractions
 are related to the bases of \cite{BR} and of this manuscript. 
We find one relationship among these contractions:
\beq
\d _{ij} \d_{kl} \d_{MN}
-  \d _{il} \d_{kj} \d_{MN}  -\d _{iN} \d_{kl} \d_{Mj} 
-\d _{ij} \d_{kN} \d_{Ml}
+  \d _{il}  \d_{kN} \d_{Mj} +  \d _{iN}  \d_{kj} \d_{Ml}
= 0~~~,
\label{dddid}
 \eeq
which will be used to remove the fourth contraction of eqn (\ref{dddcontractions}).

\subsection{Alternate bases for SU(2) doublet external fermions}

In this manuscript, we use a different  basis of dimension
eight operators from Berezhiani and Rossi,
constructed such that the operators 
 match at tree level  onto  either  NSI, or 
 LFV.

 These operators are constructed with
 doublet first generation quarks $q$ as external fermions;
 they will also be appropriate (for
 the lepton flavour indices $\{\r,\s\} \in \{ \mu,\tau\}$)
 when the external fermion is a doublet  first generation
 lepton. The dimension six operators in our basis
 are given in eqn (\ref{opsd6us}),
 and the dimension eight  operators 
 are  in eqn (\ref{opsSU2}).

Comments on this basis:
\ben
\item
${\cal O}_{NSI}$ is the
same  operator  as for singlet external fermions, and can be exchanged 
for the first contraction of eqn (\ref{dddcontractions}). It matches
at $m_W$ onto low-energy NSI.

\item The second contraction of
   eqn (\ref{dddcontractions})  is
   hermitian,  so we  exchange this $\d\d\d$ contraction for
$({\cal O}_{CCNSI} + {\cal O}^\dagger_{CCNSI}),$
which will  match at $m_W$  to NSI and CC operators.

\item
Similarly, ${\cal O}_{H2}$ is like  for external singlets,  matches
at $m_W$ only onto LFV four-fermion operators,
and corresponds to the  third contraction of eqn (\ref{dddcontractions}).

\item  the fourth  contraction of eqn (\ref{dddcontractions}) would
  match onto both NSI and LFV, so we use  the identity
  (\ref{dddid}) to remove it.  It  can be written as
 \bea
 (\overline{\ell}_\r \g_\a   \ell_\s) (\overline{q} H) \g_\a  (H^\dagger q) &=&   -  {\cal O}_{NSI} +\frac{1}{2}( {\cal O}_{CCLFV} + {\cal O}_{CCLFV}^\dagger) + \frac{1}{2}( {\cal O}_{CCNSI} + {\cal O}_{CCNSI}^\dagger)
  \eea

\item 
The last two contractions of  eqn (\ref{dddcontractions})
are 
${\cal O}_{CCLFV}$ and ${\cal O}_{CCLFV}^\dagger$,
who  match onto Charged Current and LFV operators
  below $m_W$.

The one-loop RGEs  turn out to only involve
the combination  $ C_{CCLFV,q}  + C^\dagger_{CCLFV,q}$. 
So  in the body of the manuscript, these
operators are combined into
${\cal O}_{CCLFV+} =  ({\cal O}_{CCLFV} + {\cal O}^\dagger_{CCLFV})$ .
The RGEs  are calculated
separately for $C_{CCLFV,q}^{\r\s}$,
$[C^\dagger_{CCLFV,q}]^{\r\s}$,
 $C_{CCNSI,q}^{\r\s}$,
$[C^\dagger_{CCNSI,q}]^{\r\s}$, 
then  the 
coefficient $C_+$ of ${\cal O}^{ \r\s}_{+} $
can   be  obtained by setting
\bea
C_+({\cal O} + {\cal O}^\dagger)
+
C_- ({\cal O} - {\cal O}^\dagger)
= C {\cal O} + C^\dagger{\cal O}^\dagger \;,
\nonumber
\eea
which gives  $C_+ =  (C  + C^\dagger)/2$.

\een

\section{Matching at $m_W$ }
\label{app:matching}

In this study, we should in principle use the 
one-loop minimisation condition. This is because 
the coupling constants of renormalisable interactions run,
which should be taken into account in solving the RGEs
for the operator coefficients.
If one does so,
$g, \lambda$ and $\eta$ in the anomalous
dimension matrices of eqn(\ref{solnRGEs1l})
are scale-dependent and, in  the solutions at $ \mu_f$,
should be evaluated at $\mu_f$. 
The minimisation conditions therefore should be
 expressed in terms of running parameters
at $m_W$. Then,  it is well known (see {\it eg}
 \cite{FJJ}), that   it is the sum of
the tree potential,expressed in terms of running parameters, + the  one-loop
effective potential, that is  independent
of  the renormalisation scale $\mu$.

However in practise, we often use the tree
minimisation conditions, when the RGEs give
loop contributions to LFV at the same order as
the one-loop matching conditions, because
we are only interested in the loop order at
which LFV is induced, and not in
the  precise value of the LFV operator coefficients.

It is convenient to write the one-loop  minimisation condition as
\bea
0 & = &   v \left\{ -M^2(\mu) \left( 1 +  \frac{1}{\kappa}L_{M2}\right)  +
v^2 \left( \lambda(\mu)+  \frac{1}{\kappa}L_{H2}\right) \right\} 
\equiv  v( \widetilde{M}^2 -  \widetilde{\lambda}v^2) \;.
\label{mincond}
\eea
Minimising the  one-loop effective potential given
in \cite{FJJ} (with $v^2_{here} = v^2/2|_{FJJ}$,
and  $\lambda_{here} = \lambda_{FJJ}/3$),
and evaluating at $\mu^2 = m_W^2$,  gives
\bea
L_{H2} &=& \frac{9\lambda^2}{2}
\left( \ln\frac{m_H^2}{m_W^2} -\frac{2}{3 }\right)
-6y_t^4
\left( \ln\frac{m_t^2}{m_W^2} -\frac{1}{2 }\right)
+ \frac{g^4}{8}
+ \frac{3(g^2 + g^{'2})^2}{8}
\left( \ln\frac{m_Z^2}{m_W^2} +\frac{1}{6 }\right)
\\
L_{M2} &=& \frac{3\lambda}{2}
\left( \ln\frac{m_H^2}{m_W^2} -1 \right) \;.
\eea


\begin{thebibliography}{222222}

\bibitem{NSI1} 
 L.~Wolfenstein,
  ``Neutrino Oscillations in Matter,''
  Phys.\ Rev.\ D {\bf 17} (1978) 2369.
  doi:10.1103/PhysRevD.17.2369

J.~W.~F.~Valle,
  ``Resonant Oscillations of Massless Neutrinos in Matter,''
  Phys.\ Lett.\ B {\bf 199} (1987) 432.
  doi:10.1016/0370-2693(87)90947-6

 M.~M.~Guzzo, A.~Masiero and S.~T.~Petcov,
  ``On the MSW effect with massless neutrinos and no mixing in the vacuum,''
  Phys.\ Lett.\ B {\bf 260} (1991) 154.
  doi:10.1016/0370-2693(91)90984-X


\bibitem{revFT}
  Y.~Farzan and M.~Tortola,
  ``Neutrino oscillations and Non-Standard Interactions,''
  Front.\ in Phys.\  {\bf 6} (2018) 10
  [arXiv:1710.09360 [hep-ph]].

\bibitem{BR}
  Z.~Berezhiani and A.~Rossi,
  ``Limits on the nonstandard interactions of neutrinos from e+ e- colliders,''
  Phys.\ Lett.\ B {\bf 535} (2002) 207
  [hep-ph/0111137].

\bibitem{DP-GRS}
  S.~Davidson, C.~Pena-Garay, N.~Rius and A.~Santamaria,
  ``Present and future bounds on nonstandard neutrino interactions,''
  JHEP {\bf 0303} (2003) 011
  doi:10.1088/1126-6708/2003/03/011
  [hep-ph/0302093].

\bibitem{Biggio:2009nt}
  C.~Biggio, M.~Blennow and E.~Fernandez-Martinez,
  ``General bounds on non-standard neutrino interactions,''
  JHEP {\bf 0908} (2009) 090
  doi:10.1088/1126-6708/2009/08/090
  [arXiv:0907.0097 [hep-ph]].
  
\bibitem{LBNL}
  P.~Coloma,
  ``Non-Standard Interactions in propagation at the Deep Underground Neutrino Experiment,''
  JHEP {\bf 1603} (2016) 016
  doi:10.1007/JHEP03(2016)016
  [arXiv:1511.06357 [hep-ph]].


  S.~Choubey, A.~Ghosh, T.~Ohlsson and D.~Tiwari,
  ``Neutrino Physics with Non-Standard Interactions at INO,''
  JHEP {\bf 1512} (2015) 126
  doi:10.1007/JHEP12(2015)126
  [arXiv:1507.02211 [hep-ph]].


 A.~de Gouvêa and K.~J.~Kelly,
  ``Non-standard Neutrino Interactions at DUNE,''
  Nucl.\ Phys.\ B {\bf 908} (2016) 318
  doi:10.1016/j.nuclphysb.2016.03.013
  [arXiv:1511.05562 [hep-ph]].

  J.~Liao, D.~Marfatia and K.~Whisnant,
  ``Nonstandard neutrino interactions at DUNE, T2HK and T2HKK,''
  JHEP {\bf 1701} (2017) 071
  doi:10.1007/JHEP01(2017)071
  [arXiv:1612.01443 [hep-ph]].


 S.~Fukasawa, M.~Ghosh and O.~Yasuda,
  ``Sensitivity of the T2HKK experiment to nonstandard interactions,''
  Phys.\ Rev.\ D {\bf 95} (2017) no.5,  055005
  doi:10.1103/PhysRevD.95.055005
  [arXiv:1611.06141 [hep-ph]].


 K.~Huitu, T.~J.~Kärkkäinen, J.~Maalampi and S.~Vihonen,
  ``Constraining the nonstandard interaction parameters in long baseline neutrino experiments,''
  Phys.\ Rev.\ D {\bf 93} (2016) no.5,  053016
  doi:10.1103/PhysRevD.93.053016
  [arXiv:1601.07730 [hep-ph]].

   J.~Kopp, M.~Lindner, T.~Ota and J.~Sato,
  ``Non-standard neutrino interactions in reactor and superbeam experiments,''
  Phys.\ Rev.\ D {\bf 77} (2008) 013007
  doi:10.1103/PhysRevD.77.013007
  [arXiv:0708.0152 [hep-ph]].

M.~Blennow, S.~Choubey, T.~Ohlsson, D.~Pramanik and S.~K.~Raut,
  ``A combined study of source, detector and matter non-standard neutrino interactions at DUNE,''
  JHEP {\bf 1608} (2016) 090
  doi:10.1007/JHEP08(2016)090
  [arXiv:1606.08851 [hep-ph]].
  
  M.~Masud and P.~Mehta,
  ``Nonstandard interactions and resolving the ordering of neutrino masses at DUNE and other long baseline experiments,''
  Phys.\ Rev.\ D {\bf 94} (2016) no.5,  053007
  doi:10.1103/PhysRevD.94.053007
  [arXiv:1606.05662 [hep-ph]].

  S.~K.~Agarwalla, S.~S.~Chatterjee and A.~Palazzo,
  ``Degeneracy between $\theta_{23}$ octant and neutrino non-standard interactions at DUNE,''
  Phys.\ Lett.\ B {\bf 762} (2016) 64
  doi:10.1016/j.physletb.2016.09.020
  [arXiv:1607.01745 [hep-ph]].

  
  K.~N.~Deepthi, S.~Goswami and N.~Nath,
  ``Challenges posed by non-standard neutrino interactions in the determination of $\delta_{CP}$ at DUNE,''
  Nucl.\ Phys.\ B {\bf 936} (2018) 91
  doi:10.1016/j.nuclphysb.2018.09.004
  [arXiv:1711.04840 [hep-ph]].

  
  \bibitem{atm}
   A.~Esmaili and A.~Y.~Smirnov,
  ``Probing Non-Standard Interaction of Neutrinos with IceCube and DeepCore,''
  JHEP {\bf 1306} (2013) 026
  doi:10.1007/JHEP06(2013)026
  [arXiv:1304.1042 [hep-ph]].

 T.~Ohlsson, H.~Zhang and S.~Zhou,
  ``Effects of nonstandard neutrino interactions at PINGU,''
  Phys.\ Rev.\ D {\bf 88} (2013) no.1,  013001
  doi:10.1103/PhysRevD.88.013001
  [arXiv:1303.6130 [hep-ph]].

 S.~Fukasawa and O.~Yasuda,
  ``Constraints on the Nonstandard Interaction in Propagation from Atmospheric Neutrinos,''
  Adv.\ High Energy Phys.\  {\bf 2015} (2015) 820941
  doi:10.1155/2015/820941
  [arXiv:1503.08056 [hep-ph]].
  
 S.~Fukasawa and O.~Yasuda,
  ``The possibility to observe the non-standard interaction by the Hyperkamiokande atmospheric neutrino experiment,''
  Nucl.\ Phys.\ B {\bf 914} (2017) 99
  doi:10.1016/j.nuclphysb.2016.11.004
  [arXiv:1608.05897 [hep-ph]].

 I.~Mocioiu and W.~Wright,
  ``Non-standard neutrino interactions in the mu–tau sector,''
  Nucl.\ Phys.\ B {\bf 893} (2015) 376
  doi:10.1016/j.nuclphysb.2015.02.016
  [arXiv:1410.6193 [hep-ph]].
  
\bibitem{GGetal18}
  I.~Esteban, M.~C.~Gonzalez-Garcia, M.~Maltoni, I.~Martinez-Soler and J.~Salvado,
  ``Updated Constraints on Non-Standard Interactions from Global Analysis of Oscillation Data,''
  JHEP {\bf 1808} (2018) 180
  doi:10.1007/JHEP08(2018)180
  [arXiv:1805.04530 [hep-ph]].

\bibitem{Giunti:2019xpr}
  C.~Giunti,
  ``General COHERENT Constraints on Neutrino Non-Standard Interactions,''
  arXiv:1909.00466 [hep-ph].
  
 \bibitem{SN1}
  C.~J.~Stapleford, D.~J.~Väänänen, J.~P.~Kneller, G.~C.~McLaughlin and B.~T.~Shapiro,
  ``Nonstandard Neutrino Interactions in Supernovae,''
  Phys.\ Rev.\ D {\bf 94} (2016) no.9,  093007
  doi:10.1103/PhysRevD.94.093007
  [arXiv:1605.04903 [hep-ph]].

 M.~Blennow, A.~Mirizzi and P.~D.~Serpico,
  ``Nonstandard neutrino-neutrino refractive effects in dense neutrino gases,''
  Phys.\ Rev.\ D {\bf 78} (2008) 113004
  doi:10.1103/PhysRevD.78.113004
  [arXiv:0810.2297 [hep-ph]].


\bibitem{NS}
  A.~Chatelain and M.~C.~Volpe,
  ``Neutrino propagation in binary neutron star mergers in presence of nonstandard interactions,''
  Phys.\ Rev.\ D {\bf 97} (2018) no.2,  023014
  doi:10.1103/PhysRevD.97.023014
  [arXiv:1710.11518 [hep-ph]].

  
\bibitem{deSalasSergio}
  P.~F.~de Salas and S.~Pastor,
  ``Relic neutrino decoupling with flavour oscillations revisited,''
  JCAP {\bf 1607} (2016) no.07,  051
  doi:10.1088/1475-7516/2016/07/051
  [arXiv:1606.06986 [hep-ph]].

  \bibitem{Serpico}
 P.~D.~Serpico,
  ``Standard and non-standard primordial neutrinos,''
  Phys.\ Scripta T {\bf 127} (2006) 95
  doi:10.1088/0031-8949/2006/T127/032
  [astro-ph/0606044].
  
  
\bibitem{confusion}
  P.~Huber, T.~Schwetz and J.~W.~F.~Valle,
  ``Confusing nonstandard neutrino interactions with oscillations at a neutrino factory,''
  Phys.\ Rev.\ D {\bf 66} (2002) 013006
  doi:10.1103/PhysRevD.66.013006
  [hep-ph/0202048].

  
\bibitem{ASdRR}
  D.~Aristizabal Sierra, V.~De Romeri and N.~Rojas,
  ``COHERENT analysis of neutrino generalized interactions,''
  Phys.\ Rev.\ D {\bf 98} (2018) 075018
  [arXiv:1806.07424 [hep-ph]].




\bibitem{ATZ}
  W.~Altmannshofer, M.~Tammaro and J.~Zupan,
  ``Non-standard neutrino interactions and low energy experiments,''
  arXiv:1812.02778 [hep-ph].



\bibitem{FGAX}
  A.~Falkowski, M.~González-Alonso and Z.~Tabrizi,
  ``Reactor neutrino oscillations as constraints on Effective Field Theory,''
  arXiv:1901.04553 [hep-ph].

  \bibitem{Bischer:2019ttk}
  I.~Bischer and W.~Rodejohann,
  ``General Neutrino Interactions from an Effective Field Theory Perspective,''
  Nucl.\ Phys.\ B {\bf 947} (2019) 114746
  doi:10.1016/j.nuclphysb.2019.114746
  [arXiv:1905.08699 [hep-ph]].

\bibitem{COHERENT}
  D.~Akimov {\it et al.} [COHERENT Collaboration],
  ``Observation of Coherent Elastic Neutrino-Nucleus Scattering,''
  Science {\bf 357} (2017) no.6356,  1123
  [arXiv:1708.01294 [nucl-ex]].

 D.~Akimov {\it et al.} [COHERENT Collaboration],
  ``COHERENT Collaboration data release from the first observation of coherent elastic neutrino-nucleus scattering,''
  arXiv:1804.09459 [nucl-ex].

  
\bibitem{BGN}
  S.~Bergmann, Y.~Grossman and E.~Nardi,
  ``Neutrino propagation in matter with general interactions,''
  Phys.\ Rev.\ D {\bf 60} (1999) 093008
  doi:10.1103/PhysRevD.60.093008
  [hep-ph/9903517].

\bibitem{Cirelli}
  M.~Cirelli, E.~Del Nobile and P.~Panci,
  ``Tools for model-independent bounds in direct dark matter searches,''
  JCAP {\bf 1310} (2013) 019
  doi:10.1088/1475-7516/2013/10/019
  [arXiv:1307.5955 [hep-ph]].

\bibitem{CDK}
  V.~Cirigliano, S.~Davidson and Y.~Kuno,
  ``Spin-dependent $\mu \to e$ conversion,''
  Phys.\ Lett.\ B {\bf 771} (2017) 242
  doi:10.1016/j.physletb.2017.05.053
  [arXiv:1703.02057 [hep-ph]].
  

\bibitem{polonais}
  B.~Grzadkowski, M.~Iskrzynski, M.~Misiak and J.~Rosiek,
  ``Dimension-Six Terms in the Standard Model Lagrangian,''
  JHEP {\bf 1010} (2010) 085
  [arXiv:1008.4884 [hep-ph]].


\bibitem{JMT}
  R.~Alonso, E.~E.~Jenkins, A.~V.~Manohar and M.~Trott,
  ``Renormalization Group Evolution of the Standard Model Dimension Six Operators III: Gauge Coupling Dependence and Phenomenology,''
  JHEP {\bf 1404} (2014) 159
  [arXiv:1312.2014 [hep-ph]].

  E.~E.~Jenkins, A.~V.~Manohar and M.~Trott,
  ``Renormalization Group Evolution of the Standard Model Dimension Six Operators II: Yukawa Dependence,''
  JHEP {\bf 1401} (2014) 035
  [arXiv:1310.4838 [hep-ph]].


  E.~E.~Jenkins, A.~V.~Manohar and M.~Trott,
  ``Renormalization Group Evolution of the Standard Model Dimension Six Operators I: Formalism and lambda Dependence,''
  JHEP {\bf 1310} (2013) 087
  [arXiv:1308.2627 [hep-ph]].

 
\bibitem{PSI}
  A.~Crivellin, S.~Davidson, G.~M.~Pruna and A.~Signer,
  ``Renormalisation-group improved analysis of $\mu\to e$ processes in a systematic effective-field-theory approach,''
  JHEP {\bf 1705} (2017) 117
  [arXiv:1702.03020 [hep-ph]].


\bibitem{BabuEtal}
 K.~S.~Babu, P.~S.~B.~Dev, S.~Jana and A.~Thapa,
  ``Non-Standard Interactions in Radiative Neutrino Mass Models,''
  arXiv:1907.09498 [hep-ph].
  
  
\bibitem{GXOW}
  M.~B.~Gavela, D.~Hernandez, T.~Ota and W.~Winter,
  ``Large gauge invariant non-standard neutrino interactions,''
  Phys.\ Rev.\ D {\bf 79} (2009) 013007
  [arXiv:0809.3451 [hep-ph]].



\bibitem{Antusch}
  S.~Antusch, J.~P.~Baumann and E.~Fernandez-Martinez,
  ``Non-Standard Neutrino Interactions with Matter from Physics Beyond the Standard Model,''
  Nucl.\ Phys.\ B {\bf 810} (2009) 369
  doi:10.1016/j.nuclphysb.2008.11.018
  [arXiv:0807.1003 [hep-ph]].


  
\bibitem{PP}
  M.~Pospelov and J.~Pradler,
  ``Elastic scattering signals of solar neutrinos with enhanced baryonic currents,''
  Phys.\ Rev.\ D {\bf 85} (2012) 113016
   Erratum: [Phys.\ Rev.\ D {\bf 88} (2013) no.3,  039904]
  doi:10.1103/PhysRevD.85.113016, 10.1103/PhysRevD.88.039904
  [arXiv:1203.0545 [hep-ph]].


\bibitem{FarzanModel}
  Y.~Farzan,
  ``A model for large non-standard interactions of neutrinos leading to the LMA-Dark solution,''
  Phys.\ Lett.\ B {\bf 748} (2015) 311
  doi:10.1016/j.physletb.2015.07.015
  [arXiv:1505.06906 [hep-ph]].

 \bibitem{FarzanModel2} 
  Y.~Farzan and J.~Heeck,
  ``Neutrinophilic nonstandard interactions,'' 
Phys.\ Rev.\ D {\bf 94} (2016) no.5,  053010 
doi:10.1103/PhysRevD.94.053010 
[arXiv:1607.07616 [hep-ph]]. 



\bibitem{Biggio}
  C.~Biggio, M.~Blennow and E.~Fernandez-Martinez,
  ``Loop bounds on non-standard neutrino interactions,''
  JHEP {\bf 0903} (2009) 139
  doi:10.1088/1126-6708/2009/03/139
  [arXiv:0902.0607 [hep-ph]].

\bibitem{Bellgardt:1987du}
  U.~Bellgardt {\it et al.} [SINDRUM Collaboration],
  ``Search for the Decay $\mu^+ \to e^+ e^+ e^-$,''
  Nucl.\ Phys.\ B {\bf 299} (1988) 1.

\bibitem{Bertl:2006up}
  W.~H.~Bertl {\it et al.} [SINDRUM II Collaboration],
  ``A Search for muon to electron conversion in muonic gold,''
  Eur.\ Phys.\ J.\ C {\bf 47} (2006) 337.
  doi:10.1140/epjc/s2006-02582-x
C.~Dohmen {\it et al.} [SINDRUM II Collaboration],
  ``Test of lepton flavor conservation in $\mu \to e$ conversion on titanium,''
  Phys.\ Lett.\ B {\bf 317} (1993) 631.
 W.~Honecker {\it et al.} [SINDRUM II Collaboration],
  ``Improved limit on the branching ratio $\mu \to e$ conversion on lead,''
  Phys.\ Rev.\ Lett.\  {\bf 76} (1996) 200.
  doi:10.1103/PhysRevLett.76.200

 
  
  
\bibitem{COMET}
  Y.~Kuno [COMET Collaboration],
  ``A search for muon-to-electron conversion at J-PARC: The COMET experiment,''
  PTEP {\bf 2013} (2013) 022C01.
  doi:10.1093/ptep/pts089


\bibitem{Mu2e}
 R.~M.~Carey {\it et al.} [Mu2e Collaboration],
  ``Proposal to search for $\mu^- N \to e^- N$ with a single event sensitivity below $10^{-16}$,''
  FERMILAB-PROPOSAL-0973.

\bibitem{Mu3e}
  A.~Blondel {\it et al.},
  ``Research Proposal for an Experiment to Search for the Decay $\mu \to eee$,''
  arXiv:1301.6113 [physics.ins-det].

  \bibitem{dim52}
 S.~Davidson, M.~Gorbahn and M.~Leak,
  ``Majorana neutrino masses in the renormalization group equations for lepton flavor violation,''
  Phys.\ Rev.\ D {\bf 98} (2018) no.9,  095014
  doi:10.1103/PhysRevD.98.095014
  [arXiv:1807.04283 [hep-ph]].


  
\bibitem{tmp}
  B.~Aubert {\it et al.} [BaBar Collaboration],
  ``Search for Lepton Flavor Violating Decays $\tau^\pm \to \ell^\pm \pi^0$, $\ell^\pm \eta$, $\ell^\pm \eta^\prime$,''
  Phys.\ Rev.\ Lett.\  {\bf 98} (2007) 061803
  doi:10.1103/PhysRevLett.98.061803
  [hep-ex/0610067].
  
\bibitem{tep}
  Y.~Miyazaki {\it et al.} [Belle Collaboration],
  ``Search for lepton flavor violating tau- decays into l- eta, l- eta-prime and l- pi0,''
  Phys.\ Lett.\ B {\bf 648} (2007) 341
  doi:10.1016/j.physletb.2007.03.027
  [hep-ex/0703009 [HEP-EX]].


  
\bibitem{tlr}
  Y.~Miyazaki {\it et al.} [Belle Collaboration],
  ``Search for Lepton-Flavor-Violating tau Decays into a Lepton and a Vector Meson,''
  Phys.\ Lett.\ B {\bf 699} (2011) 251
  doi:10.1016/j.physletb.2011.04.011
  [arXiv:1101.0755 [hep-ex]].

  
\bibitem{BHHS}
 D.~Black, T.~Han, H.~J.~He and M.~Sher,
  ``tau - mu flavor violation as a probe of the scale of new physics,''
  Phys.\ Rev.\ D {\bf 66} (2002) 053002
  doi:10.1103/PhysRevD.66.053002
  [hep-ph/0206056].
  
\bibitem{PDB}
``2017 Review of Particle Physics'',
C. Patrignani {\it et al.} (Particle Data Group), Chin. Phys. C, 40, 100001 (2016)
and 2017 update. 
  

\bibitem{Belletnpp}
  M.~Fujikawa {\it et al.} [Belle Collaboration],
  ``High-Statistics Study of the tau- ---> pi- pi0 nu(tau) Decay,''
  Phys.\ Rev.\ D {\bf 78} (2008) 072006
  doi:10.1103/PhysRevD.78.072006
  [arXiv:0805.3773 [hep-ex]].



\bibitem{CPX}
  A.~Celis, V.~Cirigliano and E.~Passemar,
  ``Lepton flavor violation in the Higgs sector and the role of hadronic $\tau$-lepton decays,''
  Phys.\ Rev.\ D {\bf 89} (2014) 013008
  doi:10.1103/PhysRevD.89.013008
  [arXiv:1309.3564 [hep-ph]].

\bibitem{Kou:2018nap}
  E.~Kou {\it et al.} [Belle-II Collaboration],
  ``The Belle II Physics Book,''
  arXiv:1808.10567 [hep-ex].
  
  
\bibitem{feta} 
  T.~Feldmann,
  ``Quark structure of pseudoscalar mesons,''
  Int.\ J.\ Mod.\ Phys.\ A {\bf 15} (2000) 159
  doi:10.1142/S0217751X00000082
  [hep-ph/9907491].
 H.~Leutwyler,
  ``On the 1/N expansion in chiral perturbation theory,''
  Nucl.\ Phys.\ Proc.\ Suppl.\  {\bf 64} (1998) 223
  doi:10.1016/S0920-5632(97)01065-7
  [hep-ph/9709408].
 

  

  
\bibitem{TheMEG:2016wtm}
  A.~M.~Baldini {\it et al.} [MEG Collaboration],
  ``Search for the lepton flavour violating decay $\mu ^+ \rightarrow e^+ \gamma $ with the full dataset of the MEG experiment,''
  Eur.\ Phys.\ J.\ C {\bf 76} (2016) no.8,  434
  [arXiv:1605.05081 [hep-ex]].
  
\bibitem{FJJ}
  C.~Ford, I.~Jack and D.~R.~T.~Jones,
  ``The Standard model effective potential at two loops,''
  Nucl.\ Phys.\ B {\bf 387} (1992) 373
   Erratum: [Nucl.\ Phys.\ B {\bf 504} (1997) 551]
  doi:10.1016/0550-3213(92)90165-8, 10.1016/S0550-3213(97)00532-4
  [hep-ph/0111190].



  
\end{thebibliography}
\end{document}